\documentclass[preprint]{aastex}

\newcommand{\ddt}[1]{{\partial\over\partial t}#1}
\newcommand{\grad}{{\bf \nabla}}
\renewcommand{\div}{\nabla\cdot}

\newcommand{\del}{\nabla^2}

\newcommand{\vel}{{\bf v}}
\newcommand{\pos}{{\bf x}}

\newcommand{\adota}{{{\dot{a}\over a}}}

\newcommand{\rhogas}{{\rho_{\rm g}}}
\newcommand{\rhodm}{{\rho_{\rm dm}}}
\newcommand{\velgas}{\vel_{\rm g}}
\newcommand{\veldm}{\vel_{\rm dm}}
\newcommand{\posdm}{\pos_{\rm dm}}

\newcommand{\etal}{et al.}
\newcommand{\lsim}{\mathrel{<\kern-1.0em\lower0.9ex\hbox{$\sim$}}}
\newcommand{\gsim}{\mathrel{>\kern-1.0em\lower0.9ex\hbox{$\sim$}}}

\begin{document}



\title{COSMOS: A Hybrid $N$-Body/Hydrodynamics Code for Cosmological Problems}

\author{P. M. Ricker\altaffilmark{1}}
\email{ricker@flash.uchicago.edu}

\author{S. Dodelson\altaffilmark{2}}
\email{dodelson@fnal.gov}

\author{D. Q. Lamb\altaffilmark{1}}
\email{lamb@oddjob.uchicago.edu}

\altaffiltext{1}{Department of Astronomy \& Astrophysics, University of Chicago,
       Chicago, IL 60637}
\altaffiltext{2}{Fermi National Accelerator Laboratory, Batavia, IL 60510}



\begin{abstract}

We describe a new hybrid $N$-body/hydrodynamical code based on the
particle-mesh (PM) method and the piecewise-parabolic method (PPM) for
use in solving problems related to the evolution of large-scale structure,
galaxy clusters, and individual galaxies.
The code, named COSMOS, possesses several
new features which distinguish it from other PM-PPM codes.
In particular, to solve the Poisson equation we have written a new
multigrid solver which can determine the gravitational potential of
isolated matter distributions and which properly
takes into account the finite-volume
discretization required by PPM. All components of the code are constructed
to work with a nonuniform mesh, preserving second-order spatial differences.
The PPM code uses vacuum boundary conditions
for isolated problems, preventing inflows when appropriate.
The PM code uses a second-order variable-timestep time integration scheme.
Radiative cooling and cosmological expansion terms are included.
COSMOS has been implemented for parallel computers using the
Parallel Virtual Machine (PVM) library, and it features a modular design
which simplifies the addition of new physics and the configuration of the
code for different types of problems.
We discuss the equations solved by COSMOS and describe the algorithms used,
with emphasis on these features. We also discuss the results of tests
we have performed to establish that COSMOS works and to determine its range
of validity.

\end{abstract}




\newpage


\section{Introduction}

The study of many problems of interest in astrophysics today benefits from
or even requires the use of three-dimensional hydrodynamical or $N$-body
simulations because the complex physical processes or the lack of symmetry
involved prohibit analytic or reduced-dimensionality approaches.
These problems span the entire range of astrophysical length
scales, from black hole collisions and core-collapse supernovae to clusters
of galaxies and the Lyman $\alpha$ forest.
The advent over the past two decades of modern shock-capturing
hydrodynamical algorithms and fast $N$-body methods, together with rapid
advances in available computing power, has made such
simulations feasible.

In these simulations one must make a distinction between collisional matter,
in which the mean free path between particle collisions is much smaller than
all length scales of interest, and collisionless matter, in which particles
may free-stream between collisions on length scales important to the
simulation. For collisional matter, a fluid description incorporating a
pressure field and a single-valued bulk velocity field is appropriate;
shock waves can convert energy irreversibly from bulk kinetic form to
internal form.
For collisionless matter, the bulk velocity can take on
multiple values at each point in space, and shock waves are not possible.

For simulations of galaxies, clusters of galaxies, and large-scale structure
it is important to track both kinds of matter. On the length and time scales of
interest in these simulations, baryonic matter (the intracluster or
intergalactic medium) behaves as a collisional fluid. Although locally this
gas may be partially or fully ionized, with collisions between ions and
electrons occurring by means of randomly oriented magnetic fields, on
resolvable length scales we may assume charge neutrality
and ignore local deviations
from fluid behavior (e.g., plasma instabilities).
However, because we do not yet know the identity of the dark matter,
we adopt the simplest hypothesis and assume that it is collisionless.
The
gravitational stability of apparently bound systems (galaxies and clusters of
galaxies) in which baryons provide insufficient mass
is still our best evidence for the existence of dark matter.
Thus the simplest hypothesis also treats
the interaction between these types of matter as purely gravitational.
Cosmological simulations must track the evolution of both matter constituents
in their mutual gravitational field.

Several codes now exist to simulate the combined evolution of collisional and
collisionless matter in a cosmological setting. All treat the
dark matter using superparticles which sample the particle distribution.
These $N$-body codes
differ in the methods they use to compute interparticle forces and to solve
the hydrodynamical equations for the gas. Methods for computing interparticle
forces include direct summation (PP, or particle-particle),
tree algorithms (Hernquist 1987),
particle-mesh methods (PM: Hockney \& Eastwood 1988),
and combinations of direct summation
and particle-mesh (P$^3$M: Efstathiou \& Eastwood 1981).
The tree and particle-mesh algorithms trade off force accuracy or
spatial resolution for speed in comparison with direct summation,
while variants of P$^3$M attempt to overcome the resolution limitations of
PM techniques while retaining their speed.

The hydrodynamic solvers currently in use with particle codes include variants of
smoothed-particle hydrodynamics (SPH: Gingold \& Monaghan 1977; Lucy 1977)
and grid-based methods such as the
piecewise-parabolic method (PPM: Colella \& Woodward 1984) and the
total-variation-diminishing method (TVD: Harten 1983).
SPH algorithms descended from astrophysical $N$-body
techniques and use particles to approximate the behavior of the gas, treating
gas particles as moving interpolation centers for quantities such as the gas
pressure.
Typically SPH codes achieve good spatial resolution in high-density
regions but handle shocks and low-density regions poorly.
Examples of cosmological hydrodynamics codes based on SPH include those of
Evrard (1988), Hernquist \& Katz (1989), Navarro \& White (1993),
Couchman, Thomas, \& Pierce (1995),
Steinmetz (1996),
and Owen \etal\ (1998).
Grid-based methods, which are used widely in other areas of
physics and in engineering, suffer from more limited resolution, but they
handle high-density and low-density regions equally well, and with
modern algorithms they handle shocks extremely well. Moreover, it is more
straightforward to add extra gas physics to grid-based codes.
Examples of grid-based cosmological hydrodynamics codes (excluding those
based on PPM, which are discussed below) include
the centered-difference code of Cen (1992),
the TVD code of Ryu \etal\ (1993),
the softened Lagrangian code of Gnedin (1995),
and the moving-mesh TVD code of Pen (1998).
Comparisons of various cosmological particle- and grid-based codes
have been performed by Kang \etal\ (1994) and Frenk \etal\ (1999).

The combination of PM for dark matter and PPM for gas has proven
to be an especially useful method for cosmological simulations. Accurate
shock handling and straightforward implementation of subgrid physics
argue for the use of a grid-based scheme for the gas.
When using a grid for the gas, the most efficient means for obtaining
the forces on the dark matter particles is the particle-mesh method, because
the gas and dark matter both make use of the potential defined on the grid.
Also, because the gas grid limits the spatial resolution, the greater
dynamic range of a tree code or direct
summation solver for the dark matter is not as useful.
In recent years several groups have independently developed PM-PPM codes
to study large-scale structure and galaxy cluster formation
(Bryan \etal\ 1995; Sornborger \etal\ 1996;
Gheller, Pantano, \& Moscardini 1998; Quilis, Ib\'a\~nez, \& S\'aez 1996, 98)
and cluster evolution (Roettiger, Stone, \& Mushotzky 1997). These
codes differ in a number of ways. The Bryan \etal, Sornborger \etal, and
Roettiger \etal\ codes use the Lagrange-plus-remap formulation of PPM, whereas
the Gheller \etal\ code uses the direct Eulerian formulation. Little difference
in numerical accuracy between the two formulations has been observed, but
the Lagrange-plus-remap formulation generalizes somewhat more readily
to moving or adaptive grids.
The Bryan \etal\ code
adds certain ad hoc modifications to the basic PPM scheme to improve
the resolution of narrow density peaks, and more recently this code has been
extended to include adaptive mesh refinement (AMR: Norman \& Bryan 1999).
The Quilis \etal\ code makes use of a linearized Riemann solver
due to Roe (1981), whereas the other codes use variants of the iterative
Riemann solver due to van Leer (1979).
All of the codes except that of Roettiger \etal\ use a Green's function
technique (via the Fast Fourier Transform, or FFT)
to solve for the combined gravitational potential
of the gas and dark matter. The Roettiger \etal\ code uses a conjugate
gradient solver (Hockney \& Eastwood 1988) for the Poisson equation.
To our knowledge, all of the codes use cloud-in-cell (CIC)
weighting together with a variable-timestep time integration scheme for the
particle-mesh code.

We have developed a new PM-PPM code, called COSMOS, for simulations of
large-scale structure formation and galaxy cluster evolution. Like
Gheller \etal\ (1998), we use the direct Eulerian formulation of PPM, and like
all of the above codes we use CIC weighting and a variable timestep in our
PM code. However, our code uses
a full multigrid solver rather than an FFT to obtain the gravitational
potential, enabling us to handle nonuniform grids with isolated boundary
conditions. Also, we have implemented a nonideal
equation of state following Colella \& Glaz (1985) (of use, for example, in
Ly~$\alpha$ cloud and supernova simulations), and we implement
cooling and multifluid source terms, as well as cosmological expansion
terms, with a method similar to that used
by Kritsuk, B\"ohringer, \& M\"uller (1998).
The hydrodynamical and gravitational components of COSMOS have been used
previously by Ricker (1998) to study purely hydrodynamic cluster evolution,
and the full code has been used by Ricker \& Sarazin (2000) to simulate
cluster mergers including gas and dark matter.

In this paper we describe the COSMOS code and the results of our
validation tests. Section \ref{Sec:Equations} discusses the
physics included in the code and the corresponding equations.
In Section \ref{Sec:Numerical methods} we discuss the discretization
scheme used for the grid-based components of the code, the
algorithms we use, and the modifications we have made to the standard
algorithms to suit the requirements of cosmological problems.
Section \ref{Sec:Code tests} describes the test problems we have used
to validate the code. Finally, Section \ref{Sec:Conclusion} summarizes the
paper.


\section{Equations}
\label{Sec:Equations}

In this section we describe the definitions and equations
used in the COSMOS code. All calculations take place in a three-dimensional
computational volume in which positions are specified using comoving
coordinates $\pos = {\bf r}/a$, where ${\bf r}$ is a proper position
vector and $a(t)$ is the time-dependent cosmological scale factor.
Both dark matter and gas peculiar velocities $\vel$ are considered to be
comoving; thus $\vel = \dot{\pos}$ for a particle at location $\pos$.
We define the comoving gas density and pressure as
\begin{eqnarray}
\rhogas &=& a^3\tilde{\rho}_{\rm g}\\
\nonumber
p &=& a\tilde{p}\ ,
\end{eqnarray}
\noindent where $\tilde{\rho}_{\rm g}$ and $\tilde{p}$ are the proper gas
density
and pressure, respectively. (We use the subscripts g and dm to distinguish
quantities which apply separately to both the gas and dark matter.)
Note that our definition of $p$ requires that
the comoving internal energy density be given by
\begin{equation}
\rhogas\epsilon = a\tilde{\rho}_{\rm g}\tilde{\epsilon}\ ,
\end{equation}
\noindent where $\tilde{\epsilon}$ is the proper specific internal energy
of the gas. For an ideal-gas equation of state with ratio of specific heats
$\gamma$ and average atomic mass $\mu$, we have
\begin{equation} \label{Eqn:eos}
\rhogas\epsilon = {p\over\gamma-1} = {\rhogas kT\over(\gamma-1)\mu}\ ,
\end{equation}
\noindent where
\begin{equation}
T = {\tilde{T}\over a^2}
\end{equation}
\noindent is the comoving temperature, and $\tilde{T}$ is the proper
temperature. We also define the comoving potential $\phi$ as the
solution at each simulation timestep to Poisson's equation,
\begin{equation} \label{Eqn:Poisson}
\del\phi = {4\pi G\over a^3}\left[\left(\rhogas+\rhodm\right) -
\overline{\left(\rhogas+\rhodm\right)}\right]\ .
\end{equation}
\noindent Here $\rhodm$ is the comoving density in dark matter
particles, and the bar indicates a spatial average.
The comoving total energy density of the gas is defined as
\begin{equation} \label{Eqn:enerdef}
\rhogas E \equiv \rhogas\epsilon + {1\over2}\rhogas v_{\rm g}^2 + \rhogas\phi\ ,
\end{equation}
\noindent where $v_{\rm g}=|\velgas|$ is the comoving speed of the gas.
Finally, for certain problems we make use of an equilibrium cooling
function, which we denote by $\Lambda(\rhogas,T)$. For example, cooling via
optically thin thermal bremsstrahlung emission, assuming a fully ionized gas,
would require $\Lambda\propto\rhogas^2T^{1/2}$.

We use these definitions to transform the inviscid Eulerian equations
of hydrodynamics into a simple comoving form which, in a static universe
($a\equiv 1$), reduces again to the standard equations with comoving
quantities equivalent to proper ones. The comoving gas equations solved
by COSMOS are as follows:
\begin{equation} \label{Eqn:continuity}
   \ddt{\rhogas} + \div\left[\rhogas\velgas\right] = 0
\end{equation}
\begin{equation} \label{Eqn:momentum}
   \ddt{\left(\rhogas\velgas\right)} + \div\left[\rhogas\velgas\velgas\right]
    + \grad p + 2\adota\rhogas\velgas + \rhogas\grad\phi = 0
\end{equation}
\begin{equation} \label{Eqn:totener}
   \ddt{\left(\rhogas E\right)} + \div\left[\left(\rhogas E + p\right)\velgas
   \right] +
   \adota\left[(3\gamma-1)\rhogas\epsilon + 2\rhogas v_{\rm g}^2\right]
   - \rhogas{\partial\phi\over\partial t} + \Lambda = 0
\end{equation}
\begin{equation} \label{Eqn:intener}
   \ddt{(\rhogas\epsilon)} + \div\left[\left(\rhogas\epsilon + 
   p\right)\velgas\right] - \velgas\cdot\grad p +
   \adota(3\gamma-1)\rhogas\epsilon + \Lambda = 0
\end{equation}
\noindent The equations are written in explicit conservation form.
With the finite-volume discretization used in COSMOS (Section
\ref{Sec:discretization}), we can difference these equations
(except for the internal energy equation) in such a way that errors
in the advection terms do not affect global conservation of mass,
momentum, and energy.

We determine which energy equation to use in updating the
pressure on the basis of the magnitude of the internal energy density.
During a simulation timestep we update the internal energy and
pressure according to the
internal energy equation (\ref{Eqn:intener}) and the equation of state
(\ref{Eqn:eos}). 
We update the total energy using the total energy equation
(\ref{Eqn:totener}).
At the end of the timestep, if in any zone the condition
\begin{equation} \label{Eqn:encriterion}
\rhogas\epsilon > 10^{-2}\max\left\{ {1\over 2}\rhogas v_{\rm g}^2,
|\rhogas\phi| \right\}
\end{equation}
\noindent is satisfied, we reset 
the internal energy density using equation (\ref{Eqn:enerdef}) 
with the updated values of
$\rhogas E$, $\rhogas$, $\velgas$, and $\phi$, then use the equation
of state to reset the pressure. 
If this criterion is
not satisfied, we instead reset 
the total energy density in that zone using equation (\ref{Eqn:enerdef}).
This procedure permits us to maintain total
energy conservation in regions where the pressure makes a significant
contribution to the total energy while avoiding large round-off errors in
the pressure where it is small. Condition (\ref{Eqn:encriterion}) is generally
satisfied except for high Mach-number flows and in `dust' calculations where
the gas pressure offers little support against gravity.
Similar methods based on energy (Bryan et al.\ 1995) and entropy
(Ryu et al.\ 1993) are used by other codes to handle such flows.

Equations (\ref{Eqn:continuity}) through (\ref{Eqn:intener}) suffice to
describe the behavior of matter that is collisional on the length scales
of interest to us. Examples include the gas in stellar atmospheres and in
clusters of galaxies. However, for collisionless matter components, such
as galaxies and dark matter in clusters, we must solve the equations of
motion for individual particles. These particles are subject only to
gravitational forces; hence a particle's comoving position $\posdm$ and
velocity $\veldm$ evolve according to
\begin{equation}
{d\posdm\over dt} = \veldm
\end{equation}
\begin{equation}
{d\veldm\over dt} + 2 {\dot a\over a} \veldm = - \nabla \phi ,
\end{equation}
the second term on the left in the velocity equation representing
the Hubble flow due to uniform expansion.


\section{Numerical methods}
\label{Sec:Numerical methods}

\subsection{Hydrodynamics}

\subsubsection{Finite-volume discretization}
\label{Sec:discretization}

We solve the hydrodynamical
equations described in Section \ref{Sec:Equations}
on a nonuniform, finite, Cartesian grid containing
$N_x\times N_y\times N_z$ cells. The center of the ($i,j,k$)-th cell
($i = 1\dots N_x$, $j = 1\dots Ny$, $k = 1\dots Nz$) is located at
${\bf x}_{ijk} = $($x_i,y_j,z_k$). The edges of this cell have
$x$-coordinates $x_{i\pm1/2} = x_i \pm
{1\over 2}\Delta x_i$ and $y$- and $z$-coordinates that are similarly defined.

We use a standard finite-volume discretization. A convenient way of
expressing this discretization is as follows.
We define a convolution
operator which averages a variable over a cell volume:
\begin{equation}
\label{Eqn:convop}
\alpha_{ijk}[f({\bf x})] \equiv {1\over\Delta x_i\,\Delta y_j\,\Delta z_k}
\int_{x-\Delta x_i/2}^{x+\Delta x_i/2}\int_{y-\Delta y_j/2}^{y+\Delta y_j/2}
\int_{z-\Delta z_k/2}^{z+\Delta z_k/2} d^3x'\,f({\bf x}')\ .
\end{equation}
\noindent Here $f$ is any scalar fluid variable (e.g.\ density or pressure).
Now let
\begin{equation}
f_{ijk}\equiv\alpha_{ijk}[f({\bf x}_{ijk})]\ ;
\end{equation}
\noindent by substituting a Taylor expansion for $f({\bf x}')$ about
${\bf x}_{ijk}$ we find that
\begin{equation}
f_{ijk} = f({\bf x}_{ijk}) + {\cal O}(\Delta^2)\ .
\end{equation}
\noindent Thus the fluid quantities
manipulated by this finite-volume technique differ from those used
in finite-difference methods by terms of order the square of the grid
spacing. These cell averages are to be distinguished from cell-wall averages,
which we define as follows for cell walls perpendicular to the $x$-axis:
\begin{equation}
\label{Eqn:wallavg}
f_{i\pm1/2,jk} \equiv {1\over\Delta y_j\,\Delta z_k}
\int_{y_j-\Delta y_j/2}^{y_j+\Delta y_j/2}
\int_{z_k-\Delta z_k/2}^{z_k+\Delta z_k/2} dy'dz'\,f(x_{i\pm1/2},y',z')\ .
\end{equation}
\noindent Similar definitions apply for the other coordinate directions.

If we write the
fluid equations (\ref{Eqn:continuity}) -- (\ref{Eqn:intener}) with
$x',y',z'$ as the
spatial variables, apply the convolution operator (\ref{Eqn:convop}) to both
sides, apply the divergence theorem, and then let $x=x_i$, $y=y_j$, and $z=z_k$,
we obtain a set of spatially discretized equations. As usual, the
continuity equation becomes
{\arraycolsep=0in
\begin{eqnarray}
\nonumber
\ddt{\rho_{{\rm g,}ijk}} + &\displaystyle\frac{1}{\Delta x_i}
\left[{(\rhogas v_{{\rm g,}x})_{i+1/2,jk} -
	(\rhogas v_{{\rm g,}x})_{i-1/2,jk}}\right] + & \\
\label{Eqn:spatdisc}
&\displaystyle\frac{1}{\Delta y_j}
\left[{(\rhogas v_{{\rm g,}y})_{i,j+1/2,k} -
	(\rhogas v_{{\rm g,}y})_{i,j-1/2,k}}\right] + & \\
\nonumber
&\displaystyle\frac{1}{\Delta z_k}
\left[{(\rhogas v_{{\rm g,}z})_{i,j,k+1/2} -
	(\rhogas v_{{\rm g,}z})_{i,j,k-1/2}}\right] = &\ 0\ .
\end{eqnarray}
}
\noindent The other fluid equations transform in a similar fashion.
In this form the advection terms in equations
(\ref{Eqn:continuity}) -- (\ref{Eqn:totener}) are conservatively
differenced, because the same amount of each
conserved quantity is subtracted from each cell as is added to its neighbors.
The nonconservative advection term in the internal energy equation
(\ref{Eqn:intener}) does not significantly affect global energy conservation,
because wherever the internal energy density makes a large
contribution to the total it is reset using the total energy density,
which is conservatively differenced.

Because the hydrodynamical
equations are hyperbolic, we solve them by integrating forward
in time from a given initial state. 
We use an explicit forward time difference, so for stability
we limit the timestep
using the Courant-Friedrichs-Lewy (CFL) criterion (Roache 1998):
\begin{equation}
\Delta t_{\rm hydro} = \sigma \min_{\rm grid}
	   \left[{ {{\Delta x_i}\over {|v_{{\rm g,}x,ijk}|+|c_{{\rm s},ijk}|}}, \,
		   {{\Delta y_j}\over {|v_{{\rm g,}y,ijk}|+|c_{{\rm s},ijk}|}}, \,
		   {{\Delta z_k}\over {|v_{{\rm g,}z,ijk}|+|c_{{\rm s},ijk}|}}
		   }\right]\ .
\end{equation}
\noindent Here $\sigma$ is the `CFL parameter', a number between 0 and 1
(as close to 1 as possible for accuracy and for computational efficiency),
and the minimization is taken over the entire computational grid.
Additional restrictions on the timestep due to gravitational acceleration
($\Delta t_{\rm grav}$), radiative cooling
($\Delta t_{\rm cool}$), and the $N$-body code ($\Delta t_{\rm dm}$)
are described in the following sections. We adopt a timestep $\Delta t$
which is the minimum of all of these restrictions:
\begin{equation}
\Delta t = \min\left[\Delta t_{\rm hydro}, \Delta t_{\rm grav},
		     \Delta t_{\rm cool}, \Delta t_{\rm dm}, \ldots\right]\ .
\label{Eqn:dt}
\end{equation}

To obtain the fully discretized equations we average the spatially
discretized equations over the time interval $(t_n,t_{n+1})$, where
$\Delta t_n = t_{n+1} - t_n$. The continuity equation then becomes
{\arraycolsep=0pt
\begin{eqnarray}
\nonumber
\rho^{n+1}_{{\rm g,}ijk} = \rho^n_{{\rm g,}ijk} &-&
  \displaystyle \frac{\Delta t_n}{\Delta x_i}
  \left[{\overline{(\rhogas v_{{\rm g,}x})}_{i+1/2,jk} -
  \overline{(\rhogas v_{{\rm g,}x})}_{i-1/2,jk}}\right] \\
  \label{Eqn:fulldisc}
&-&
  \displaystyle \frac{\Delta t_n}{\Delta y_j}
  \left[{\overline{(\rhogas v_{{\rm g,}y})}_{i,j+1/2,k} -
  \overline{(\rhogas v_{{\rm g,}y})}_{i,j-1/2,k}}\right] \\
  \nonumber
&-&
  \displaystyle \frac{\Delta t_n}{\Delta z_k}
  \left[{\overline{(\rhogas v_{{\rm g,}z})}_{i,j,k+1/2} -
  \overline{(\rhogas v_{{\rm g,}z})}_{i,j,k-1/2}}\right]\ ,
\end{eqnarray}
}
\noindent where bars indicate averages over $(t_n,t_{n+1})$.
This is an exact equation for $\rho^{n+1}_{{\rm g,}ijk}$; discretization error
is introduced when we estimate the time-averaged fluxes
($\overline{(\rhogas v_{{\rm g,}x})}_{i+1/2,jk}$, etc.)\ given the
cell-averaged
fluid variables at time $t_n$, since the fluxes depend on cell-wall
averages which must be determined through interpolation. In COSMOS
the interpolation and flux computation steps are handled using the
piecewise-parabolic method (PPM), which we discuss in the next section.

To simplify the algorithm, we reduce this three-dimensional problem to a
set of one-dimensional problems using standard operator splitting techniques.
This allows us to use
a one-dimensional hydrodynamics routine which operates on 1D work arrays, into
which and out of which we swap rows or columns from the full grid as necessary.
In the simplest form of operator splitting, one advances every cell through
$\Delta t$ using just the
$x$-derivatives, then takes the results of this operation and advances them
again through $\Delta t$ using just the $y$-derivatives, finally
repeating the procedure using just the $z$-derivatives.
This method produces an effective three-dimensional operator which is
accurate to ${\cal O}(\Delta t)$. We have found in spherical explosion tests
that first-order splitting does not preserve rotational symmetry well.
Therefore we use the symmetrized operator splitting method of
Strang (1968), which yields ${\cal O}(\Delta t^2)$ accuracy. For each
successive
pair of timesteps we perform first a sweep in $x$, then a sweep
in $y$, then a sweep in $z$, each for a full timestep.
We then repeat these operators in reverse order for the second timestep.

Splitting techniques also generalize to the other operators in the
problem. For the cosmological expansion terms we use the exact solution of
\begin{eqnarray}
\ddt{(\rhogas\velgas)} &=& -2\adota\rhogas\velgas\\
\ddt{(\rhogas E)} &=& -\adota\left[\left(3\gamma-1\right)\rhogas\epsilon +
	2\rhogas v_{\rm g}^2\right]\\
\ddt{(\rhogas\epsilon)} &=& -\adota\left(3\gamma-1\right)\rhogas\epsilon\ .
\end{eqnarray}
\noindent We solve the radiative cooling term using a semi-implicit numerical
ODE integrator. For stability both operators require timestep limitations:
for the cosmological terms, we limit the change in $a(t)$ during the timestep
to $10\%$ or less, while for the radiative cooling term we use
\begin{equation}
\Delta t_{\rm cool} = 0.25{\rhogas\epsilon\over\Lambda}\ .
\end{equation}
These timestep restrictions are considered along with the hydrodynamical
and $N$-body restrictions in determining the actual timestep via equation
(\ref{Eqn:dt}).

We also treat the gravitational acceleration terms in an operator-split
fashion, but because they involve spatial derivatives, we must make a small
modification to the PPM method to accomodate them.
This modification is described by Colella \& Woodward (1984).
In Section \ref{Sec:gravdisc} we discuss further the discretization of the
Poisson equation and the expressions used for the cell-averaged gravitational
acceleration in PPM.
We impose the following timestep constraint due to the gravitational
acceleration ${\bf g}_{ijk}$:
\begin{equation}
\Delta t_{\rm grav} = \min_{\rm grid}\left[
				{\Delta x_i\over |g_{x,ijk}|},
				{\Delta y_j\over |g_{y,ijk}|},
				{\Delta z_k\over |g_{z,ijk}|}\right]^{1/2}\ .
\end{equation}
In practice this criterion rarely determines the timestep.


\subsubsection{The piecewise-parabolic method}

To calculate the time-averaged fluxes we use 
the piecewise-parabolic method (PPM; Colella \& Woodward (1984)),
a high-order Eulerian generalization of Godunov's (1959) scheme.
Godunov schemes achieve very good resolution of flow discontinuities such
as shocks without excessive dissipation or oscillations by solving a
nonlinear flow problem at each cell interface.
This property makes PPM ideal for solving a variety of astrophysical
flow problems, since such problems often require narrow hydrodynamical
features to be resolved with a small number of cells.

Since the PPM algorithm is discussed in detail elsewhere,
we limit ourselves to describing the special features of the version
of PPM used in COSMOS. We implement PPM using the direct Eulerian
formulation, and we include the modifications developed by Colella \& Glaz
(1985)
for handling general equations of state using a variable ratio of specific
heats $\gamma$. We use the approximate Riemann solver described by
van Leer (1979); rarefactions are locally approximated as shocks, following
Colella (1985).

PPM uses information from four cells on either side of
each cell at each timestep. At the boundaries of the computational region
we therefore maintain four `ghost zones' containing boundary information.
We set the values in these zones before applying each 1D differential operator.
Periodic boundaries, which are appropriate for large-scale structure
simulations, are implemented by setting the ghost zones equal to the
corresponding interior zones on the opposite side of the grid. For galaxy
cluster evolution problems it is more desirable to control inflow and outflow
rates than to have matter which disappears from the grid reappear on the
opposite side. For inflow boundaries one sets the ghost zones to prescribed
values; for outflow, one typically uses Neumann or zero-gradient boundaries
(Roache 1998). In the latter case the interior zones adjacent
to the boundaries are simply copied into the ghost zones. For most
astrophysical outflow applications this approach is fine, although strictly
speaking when the outflow is subsonic one should use a method-of-lines-based
approach to prevent reflections from propagating into the interior
(Thompson 1987).

However, when zero-gradient outflow boundaries are used with a
self-gravitating fluid, one must take care that the flow in the interior
zones adjacent to the boundary is indeed directed outward, else the copying
of the velocity component normal to the boundary into the ghost zones
will create an
artificial (and destabilizing) inward flux. For isolated problems we
do not prescribe an inflow pattern. In such cases we therefore
implement a form of vacuum boundary in which Neumann conditions are
used unless the normal velocity in the zones just interior to a boundary is
directed inward. If the normal component of velocity is directed inward, we
use Dirichlet (zero-value) boundaries for it and Neumann conditions for the
other variables, and we zero the external fluxes (resulting from the solution
of Riemann problems on the external boundaries) if they happen to be
inwardly directed. In isolated calculations, therefore,
matter which leaves our computational grid
disappears altogether and cannot return. Since the gas is self-gravitating,
this would cause problems with the potential if a large clump of material
were to leave the grid suddenly. We must therefore take care to select a large
enough grid to prevent substantial amounts of material from leaving during
the course of a simulation.


\subsection{Gravitation}

\subsubsection{Finite-volume discretization}
\label{Sec:gravdisc}

At the end of each timestep we solve the Poisson equation (\ref{Eqn:Poisson})
for the combined gravitational potential of the gas and dark matter.
If we apply the spatial convolution operator to this equation and use Taylor
expansions about the cell walls to obtain cell-wall averages of
$\partial\phi/\partial x$, $\partial\phi/\partial y$, and $\partial\phi/
\partial z$, we obtain the
following second-order discretized version, in which only the $x$-derivatives
are presented for clarity:
\begin{eqnarray}
\label{Eqn:discpois}
\lefteqn{
\left[B^+_{2,i}\phi_{i+2,jk} + B^+_{1,i}\phi_{i+1,jk} +
(B^+_{0,i}-B^-_{0,i})\phi_{ijk} - B^-_{1,i}\phi_{i-1,jk} -
B^-_{2,i}\phi_{i-2,jk}\right] + \ldots = } \hspace{3in} \\
\nonumber
&&
{4\pi G\over a^3}
    \left[(\rho_{{\rm g,}ijk}+\rho_{{\rm dm,}ijk}) -
    \overline{\rhogas+\rhodm}\right]\ .
\end{eqnarray}
Here we define
\begin{eqnarray}
\nonumber
A^\pm_i    &\equiv& \pm \displaystyle{2\over\Delta x_i
		(\Delta x_i+\Delta x_{i\pm1}+\Delta x_{i\pm2})} \\
\label{Eqn:discpois coeffs}
B^\pm_{0,i}&\equiv& -A^\pm_i {2\Delta x_{i\pm1}+\Delta x_{i\pm2}
		\over\Delta x_i+\Delta x_{i\pm1}} \\
\nonumber
B^\pm_{1,i}&\equiv& -A^\pm_i {\Delta x_i^2 - 3\Delta x_{i\pm1}
		(\Delta x_{i\pm1}+\Delta x_{i\pm2}) -
	        \Delta x_{i\pm2}^2\over(\Delta x_i+\Delta x_{i\pm1})
		(\Delta x_{i\pm1}+\Delta x_{i\pm2})} \\
\nonumber
B^\pm_{2,i}&\equiv& A^\pm_i {\Delta x_i-\Delta x_{i\pm1}\over
		\Delta x_{i\pm1}+\Delta x_{i\pm2}}\ .
\end{eqnarray}
The spatially averaged total density $\overline{(\rhogas+\rhodm)}$
is subtracted from the total density to make the potential tend to zero
at large distances from a point source. The cell-averaged dark matter density
$\rho_{{\rm dm,}ijk}$ is provided by the particle-mesh code.
The complicated form of equations
(\ref{Eqn:discpois})--(\ref{Eqn:discpois coeffs}) is due to the nonuniform
grid. If the spacings $\Delta x_i$ are identical, the left-hand side 
of equation (\ref{Eqn:discpois}) reduces
to the usual second-order difference
$(\phi_{i+1} - 2\phi_i + \phi_{i-1})/\Delta x^2$.

The densities on the right-hand side of
equation (\ref{Eqn:discpois}) are cell-averaged quantities, so 
the potential which
results from solving this equation is also a cell-averaged quantity.
This raises the
question of how to obtain the gravitational acceleration ${\bf g}_{ijk}$,
which the discretized gas momentum and energy equations require to be
a cell-averaged quantity since it is a source
term. With the spatial convolution operator we can obtain an expression for
the $x$-component of ${\bf g}_{ijk}$:
\begin{eqnarray}
g_{x,ijk} &\equiv& -{1\over\Delta x_i\,\Delta y_j\,\Delta z_k}
 \int_{x_i-\Delta x_i/2}^{x_i+\Delta x_i/2}
 \int_{y_j-\Delta y_j/2}^{y_j+\Delta y_j/2}
 \int_{z_k-\Delta z_k/2}^{z_k+\Delta z_k/2}
 d^3x'\,{\partial\over\partial x'}\phi({\bf x}') \\
\label{Eqn:gravacc}
\nonumber
&=& -\displaystyle\frac{1}{\Delta x_i}
 \left[{\phi_{i+1/2,jk} - \phi_{i-1/2,jk}}\right]\ .
\end{eqnarray}
\noindent 
We compute the cell-wall-averaged potential
($\phi_{i+1/2,jk}$, etc.)\ from
the cell averages $\phi_{ijk}$ using Taylor expansions of
$\phi({\bf x})$ about the cell walls in each direction.
As an example, the results for $\phi_{i-1/2,jk}$ and $\phi_{i+1/2,jk}$ are
\begin{equation}
\phi_{i\pm1/2,jk} = C^\pm_{0,i}\phi_{ijk} + C^\pm_{1,i}\phi_{i\pm1,jk} + 
		    C^\pm_{2,i}\phi_{i\pm2,jk}\ ,
\end{equation}
where
\begin{eqnarray}
\nonumber
C^\pm_{0,i} &\equiv&
  {\Delta x_{i\pm1}(\Delta x_{i\pm1}+\Delta x_{i\pm2})\over
  (\Delta x_i+\Delta x_{i\pm1})(\Delta x_i+\Delta x_{i\pm1}+\Delta
  x_{i\pm2})}\\
C^\pm_{1,i} &\equiv&
  {\Delta x_i\left[\Delta x_{i\pm1}(2\Delta x_i+
  3(\Delta x_{i\pm1} + \Delta x_{i\pm2})) + \Delta x_{i\pm2}
  (\Delta x_i+\Delta x_{i\pm2})\right]\over (\Delta x_i+
  \Delta x_{i\pm1})(\Delta x_{i\pm1}+
  \Delta x_{i\pm2})(\Delta x_i+\Delta x_{i\pm1}+\Delta x_{i\pm2})}\\
\nonumber
C^\pm_{2,i} &\equiv&
  - {\Delta x_i\Delta x_{i\pm1}\over
  (\Delta x_{x\pm1}+\Delta x_{i\pm2})(\Delta x_i+\Delta x_{i\pm1}+
  \Delta x_{i\pm2})} \ .
\end{eqnarray}
For each cell we calculate
${\bf g}_{ijk}$ in this way using the cell average of $\phi$ in that
cell and
its four nearest neighbors in each coordinate direction, yielding second-order
accuracy for ${\bf g}_{ijk}$.
If the mesh spacing is uniform, the resulting expression for $g_{x,ijk}$
is
\begin{equation}
g_{x,ijk} = \displaystyle\frac{1}{6\Delta x}\left[
	\phi_{i+2,jk} - 5(\phi_{i+1,jk} - \phi_{i-1,jk}) - \phi_{i-2,jk}
	\right]\ . 
\end{equation}

\subsubsection{The multigrid solver}

We solve the Poisson equation (\ref{Eqn:Poisson}) using the full multigrid (FMG)
method (Brandt 1977; Hackbusch 1985).
This method is as fast as 
the direct transform-based methods but can be parallelized more easily
and in a more machine-independent fashion. It also generalizes more easily
to nonuniform grids and nonperiodic boundaries.
In addition to the finite-volume discretization described in the previous
section, our multigrid implementation features the capability of handling
isolated boundary conditions.
Note that our use of multigrid to
compute the gravitational potential is to
be distinguished from the solution of potential-flow problems in large-scale
structure simulations, in which the hydrodynamical
equations also are solved using multigrid techniques.

\begin{figure}
\plotone{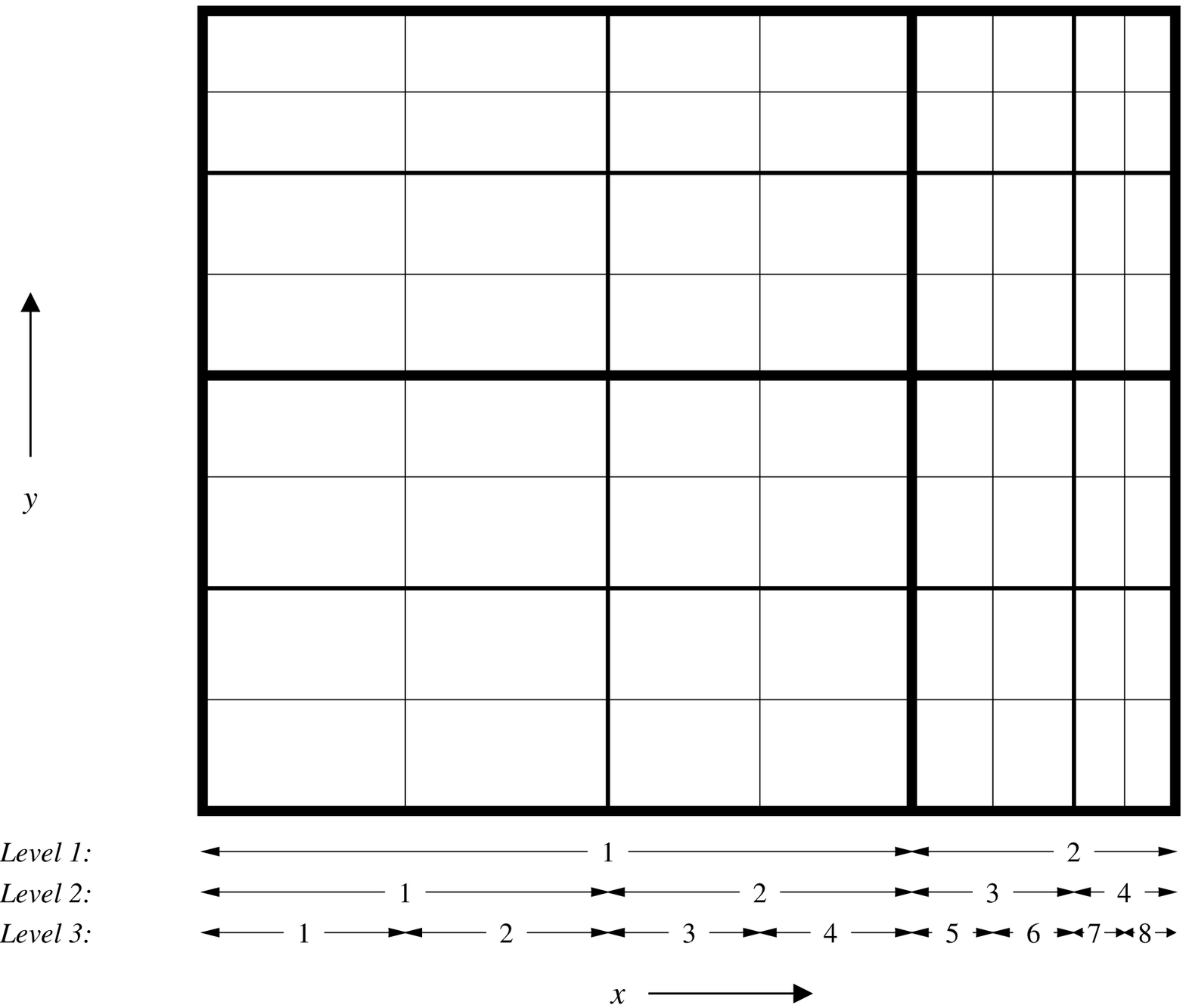}
\caption{
  Example multigrid hierarchy for a
  2D grid which is nonuniform in the $x$-direction and uses three levels of
  grid refinement. The thickness of the lines bounding a cell indicate its
  grid level; for example, the thickest lines indicate cell divisions on the
  coarsest level. The $x$-coordinates of cell centers on each level are
  indicated by numbers.
  \label{Fig:Multigrid Hierarchy}
  }
\end{figure}

To implement FMG, we begin
by constructing a hierarchy of nested grids, each of which is twice as coarse
as the previous one (Figure \ref{Fig:Multigrid Hierarchy}). 
The hierarchy starts with a grid only a few zones across, on which the
Poisson equation can be solved directly, and ends with a grid identical to
that used for the hydrodynamical equations. On each grid level FMG applies an
iterative solution method (here, Jacobi),
bringing errors on all length scales into convergence at the same rate.

\begin{figure}
\plotone{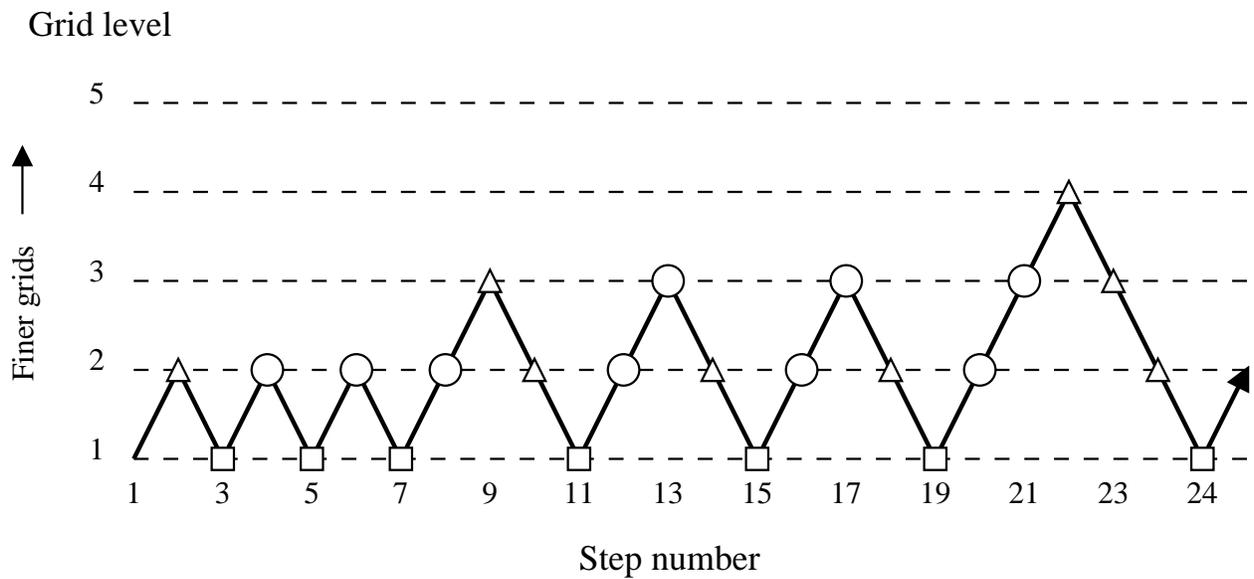}
\caption{
  Schematic diagram of steps in our
  multigrid solver. Ascending line segments indicate prolongation, and
  descending
  line segments indicate restriction. Squares indicate direct solution on the
  coarsest level, triangles indicate computation of the residual, and circles
  indicate correction of the residual.
  \label{Fig:Multigrid V-Cycle}
  }
\end{figure}

Our multigrid scheme uses an ascending V-cycle (Figure
\ref{Fig:Multigrid V-Cycle}).
We begin with a guess for the solution on the coarsest grid.  We first apply a
few Jacobi iterations to the guess (this smooths the 
error modes with wavelengths which are short relative to the grid spacing),
then `prolongate' this approximate solution to the next finer level. Our
prolongation operator is simple trilinear interpolation using cell-averaged
quantities with nonuniform (though nested) grids. After performing a few
Jacobi iterations at this level, we
then calculate the `residual', the difference between the guess and the
result of applying the Laplacian operator to the density on this grid.
Since the Poisson
equation is linear, the solution to the residual equation is just the
correction
which must be added to the original guess to obtain the true solution.
However, since long-wavelength error modes are responsible for the Jacobi
method's slow convergence, the residual on the second
level will be dominated by errors that have long wavelengths relative to
this grid and shorter wavelengths relative to the coarsest grid. To remove these
errors, then, we `restrict' the residual back to the coarsest grid and there
solve exactly for the correction (actually several Jacobi iterations on
this level are sufficient). Since we are using finite volumes, restriction
is equivalent to simple averaging over the fine-grid zones which lie within
a given coarse-grid zone. After solving on the coarsest grid for the
correction, we prolongate the correction back to the second level and apply
it to the solution guess on this level. We repeat this V-cycle twice, then
prolongate the solution guess to the next finer level, where we perform three
more V-cycles, each time restricting all the way back to the coarsest level.
We repeat this process until we reach the finest
level (the original grid).

Boundary zones for the multigrid solver must be maintained both on the
sides and at the corners of the computational cube for each
level of the grid hierarchy, although the corners are only used for
prolongations. The grid levels are nested, so their external
boundaries coincide, and Dirichlet, Neumann, or periodic boundaries
are simple to implement. The primary complication lies in distinguishing
between cell-averaged values and cell-wall-averaged values of the potential
and the density; the cell averages in the ghost zones must be chosen so as
to make the cell-wall averages of the potential on the boundary possess the
desired properties. We do this by writing the potential as a Taylor series
about the boundary location in the direction perpendicular to the boundary,
then average the series over nearby cells,
obtaining a set of equations for the cell-wall averages of the potential and
its derivatives in terms of the (known and unknown) interior and ghost-zone
cell-averaged potential. We invert these equations (up to some limiting order)
to obtain the unknown ghost-zone cell average(s) in terms of the known
boundary values and interior-zone cell averages.

Isolated boundaries are boundaries outside of which the
source function is identically zero. Implementing them requires that we
specify the value of the potential on the boundary surface.
In order to solve for the gravitational potential of isolated matter,
we use a variant of James' (1977) method.
We first calculate the potential assuming Dirichlet boundaries.
We then find the image distribution required to make the potential
identically zero outside the boundary by evaluating the Laplacian of
this Dirichlet solution on the boundary. Finally,
we calculate the isolated potential of the (hollow) image
distribution by computing its moments up to some maximum multipole
and solving with
boundary values appropriate to the corresponding multipole
expansion of the potential. By subtracting this from the Dirichlet
solution we obtain the desired isolated potential. We reduce the number
of multipole moments required by first finding the center of mass of the
image distribution and then performing the multipole expansion about
this point. Our method is somewhat similar to that of M\"uller \& Steinmetz
(1995), who use
a spherical-harmonic expansion of the original source function about the
center of a spherically symmetric grid to compute the potential at each of
the interior grid points.


\subsection{Dark matter}

To handle collisionless components, such as dark matter and stars, we use
an $N$-body code based on the particle-mesh (PM) method
(Efstathiou et al.\ 1985; Hockney \& Eastwood 1988).
This technique speeds force calculations for particles
by converting particle positions to densities on a grid, then solving the
Poisson equation on the grid using a fast direct or multigrid solver, and
finally interpolating the potential from the grid to obtain forces at the
particle locations. Since we already need to solve the Poisson equation on
a grid for our hydrodynamical code, the particle-mesh method is ideally
suited for integration with this code. We simply add the equivalent densities
for the particles to the grid densities for the gas, then solve the Poisson
equation as usual. In our case, grid densities for the particles are
computed using the cloud-in-cell (CIC) operator. We also use this operator
to compute interpolated forces.

Other $N$-body methods, including particle-particle-particle-mesh (P$^3$M)
and the tree methods, are often used to follow the dark matter in cluster
simulations. The P$^3$M method (Efstathiou \& Eastwood 1981)
extends the resolution of particle-mesh by
using direct summation for the force between particles lying within a single
zone. Tree methods
(Hernquist 1987), on the other hand, approximate the force due to distant
groups of particles using their low-order multipole moments. However, because
we are using a grid-based hydrodynamical method, we do not expect to resolve
features smaller than a single zone. Therefore the considerable
execution time expended in the direct summation part (which scales as the
square of the number of particles) of P$^3$M would be wasted. The extra
resolution provided by tree codes would likewise be wasted. Particle-mesh
has another advantage for us over tree methods in the ease with which it can
be integrated with the gas code.

The main difference between the PM part of COSMOS and the `classic' PM codes 
(e.~g., Efstathiou et al.\ 1985) is our use of a variable timestep. 
In order to keep the computation accurate to second order, we need a more
complicated version of the standard leapfrog update of position and velocity
which makes use of the gravitational acceleration on each particle stored from
the preceding timestep.
The resulting update steps (given here only for the $x$ component of position and
velocity) are
\begin{equation}
x_{n+1} = x_n + v_{n+1/2} \Delta t_n \end{equation}
for the position and
\begin{eqnarray}
v_{n+1/2} &=& v_{n-1/2} \left[ 1 - {A_n\over 2} \Delta t_n
+ {1\over 3!} \Delta t_n^2
\left( A_n^2 -\dot A_n \right)\right]
\left[ 1 - \Delta t_{n-1} {A_n\over 2}
+ \Delta t_{n-1}^2 {A_n^2+ 
2\dot A_n\over 12} \right]\nonumber
\\ 
&& - \nabla \phi_{n} \left[ {\Delta t_{n-1}\over 2} +
{\Delta t_n^2\over
6\Delta t_{n-1}} + {\Delta t_{n-1} \over 3} - { A_n \Delta t_n \over 6}
\left( \Delta t_n + \Delta t_{n-1} \right)
\right] \nonumber
\\ 
&& - \nabla \phi_{n-1} \left[ {\Delta t_{n-1}^2 - \Delta t_n^2
\over 6\Delta t_{n-1}}
- { A_n \Delta t_{n-1} \over 12} (\Delta t_n + \Delta t_{n-1}) \right]
\label{Eqn:timestep}
\end{eqnarray}
for the velocity. Here we define $A \equiv -2 \dot a/a$. For the first
half-timestep we use a
first-order accurate expression to obtain $v_{1/2}$ from the initial conditions.
Note that 
the derivatives of the potential from the previous timestep (here denoted 
$\nabla \phi_{n-1}$) must be retained. This is also necessary in the PPM
code in order to make
second-order accurate gravitational corrections to the Riemann solver,
as described by Bryan et al.\ (1995).



\section{Code tests}
\label{Sec:Code tests}

As with any scientific code of this complexity it is necessary to apply COSMOS
to a number of test problems with known solutions
before using it on research problems. This is important not simply for the
purpose of verifying that the code works as it should, but
also because it gives us an intuitive understanding of the code's
strengths and weaknesses. Such an understanding is
essential for making effective use of any hydrodynamical code. 
Accordingly, we have performed a number of tests which exercise the various code
modules in different combinations. We describe the test problems and the
code's performance on each in this section.


\subsection{Hydrodynamics tests}

\subsubsection{Sod shock-tube problem}

The widely used test problem due to Sod (1978) is a Riemann shock tube
problem with initial conditions specially chosen to produce all three types
of fluid discontinuity (shock, contact discontinuity, and rarefaction).
In this test we create two regions of constant density and pressure,
separated by a plane whose normal forms specified angles with the $x$ and
$y$ axes. To the left of the plane we set $\rho_{\rm L} = p_{\rm L} = 1$,
and to the right we set $\rho_{\rm R} = 0.125$ and $p_{\rm R} = 0.1$. The
ratio of specific heats, $\gamma$, is 1.4, and the fluid is everywhere at
rest. This test is purely hydrodynamical; no gravitational potential is used.
The Sod test enables us to determine if our code satisfies the shock jump
conditions and whether it correctly determines the speed of each nonlinear wave.
By performing this essentially one-dimensional test in three dimensions
at an angle to the grid, we can also determine how well we can resolve shocks
for realistic grid sizes in 3D.

\begin{figure}
\epsscale{0.85}
\plotone{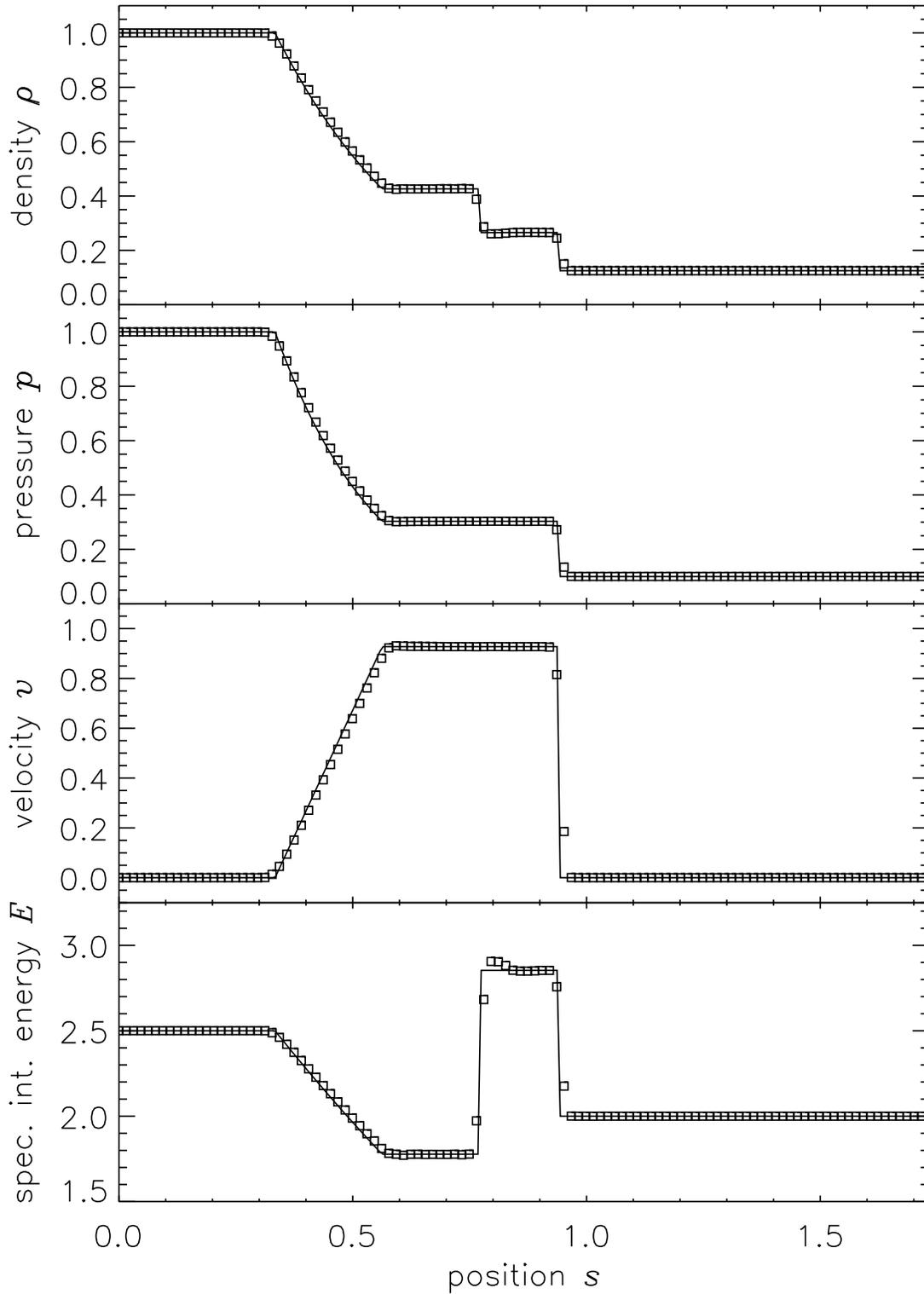}
\caption{
  Density, pressure, velocity, and specific
  internal energy in the Sod problem at $t = 0.206$.
  Solid lines represent the analytic solution; squares
  indicate numerical solution values interpolated along the line normal to
  the shock which passes through the origin.
  \label{Fig:Sod Snapshot}
  }
\end{figure}

Figure \ref{Fig:Sod Snapshot} plots our numerical solution to the Sod problem
against the analytic solution at $t=0.206$. We used a $64^3$ uniform grid
in the box $[0,1]^3$; the figure shows profiles taken as a function of distance
$s$ along the line segment connecting (0,0,0) and (1,1,1)
(normal to the shock plane).
The initial fluid discontinuity was located at $s = 0.577$.
In constructing the initial conditions we smoothed the
discontinuity in each variable over one zone's width to reduce the
starting error resulting from the presence of all three nonlinear waves
within one zone at early times. Since errors in the contact discontinuity
propagate at the speed of the fluid (and hence of the discontinuity), they
accumulate at the discontinuity, and any starting error will therefore be
present at all later times. Even with the initial smoothing a 2\% error is
present in the density in the three zones in front of the contact
discontinuity; this results in a similar error in the specific internal
energy. Apart from this error and a slight underestimation of the slopes
in the leftward-moving rarefaction wave, the numerical solution is nearly
exact. The position of each of the three discontinuities is correct, and the
shock and contact discontinuity are each resolved using about two zones,
despite the fact that the shock is moving at an angle to the grid. This
excellent shock handling is one of the primary reasons for using PPM.

\subsubsection{Sedov explosion problem}

To determine how well our hydrodynamical code respects the $90^\circ$
rotational symmetry of our grid, and to determine whether this behavior
carries over to a nonuniform grid, we studied the expansion of a 
strong spherical shock wave into a uniform medium. In this problem, which
is again purely hydrodynamical, we deposit a quantity of energy $E=1$ into a
small sphere of radius $\delta r$ at the center of the grid.
The pressure inside this volume, $p_0'$, is given by
\begin{equation}
p_0' = {3(\gamma-1)E\over 4\pi\,\delta r^3}\ ,
\end{equation}
\noindent where for this problem we use $\gamma=7/5$. Everywhere the density
is set equal to $\rho_0=1$, and everywhere but the center of the grid the
pressure is set to a small value, $p_0=10^{-5}$.
The fluid is initially at rest.
Sedov (1959) first obtained a self-similar analytic
solution to this problem. A spherically symmetric shock wave develops; the
density, pressure, and radial velocity are all functions of
\begin{equation}
\xi \equiv {r\over R(t)} = {r\over C}\left({\rho_0\over
	E t^2}\right)^{1/5}\ ,
\end{equation}
\noindent where $C$ is a numerical constant depending only on $\gamma$.
Just behind the shock front at $\xi = 1$ we have
\begin{eqnarray}
\nonumber
\rho = & \rho_1 \equiv & {\gamma+1\over\gamma-1}\rho_0 \\
p    = & p_1    \equiv & {2\over\gamma+1}\rho_0 u^2 \\
\nonumber
v    = & v_1    \equiv & {2\over\gamma+1}u\ ,
\end{eqnarray}
\noindent where $u \equiv dR/dt$ is the speed of the shock wave. Near the
center of the grid,
\begin{eqnarray}
\nonumber
\rho(\xi)/\rho_1 & \propto & \xi^{3/(\gamma-1)} \\
p(\xi)/p_1       & =       & {\rm constant} \\
\nonumber
v(\xi)/v_1       & \propto & \xi\ .
\end{eqnarray} 

\begin{figure}
\epsscale{0.85}
\plotone{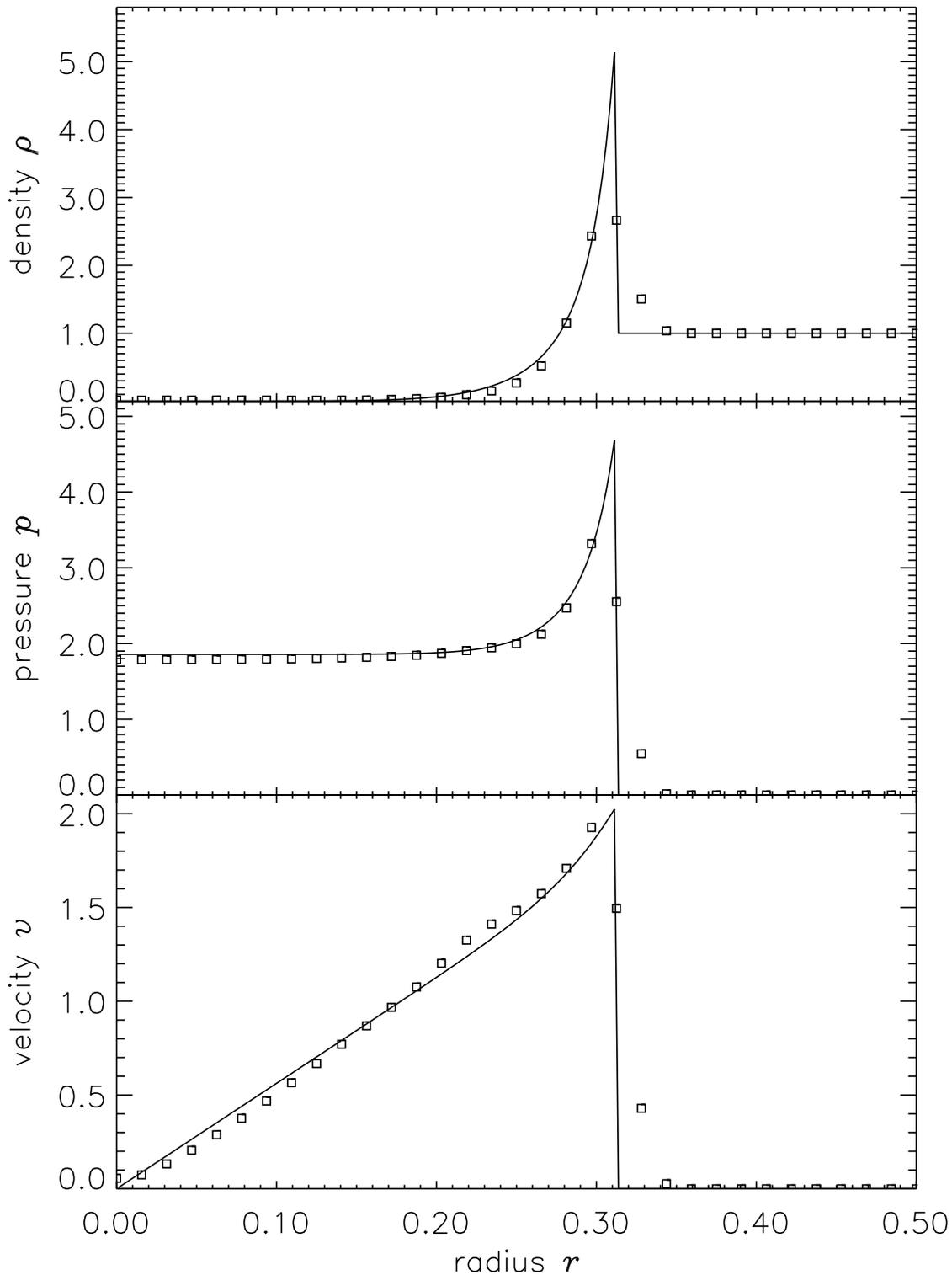}
\caption{
  Density, pressure, and velocity
  for the Sedov explosion problem on a uniform $64^3$ grid at $t=0.0508$.
  Squares represent the angle-averaged numerical solution, and
  solid lines represent the analytic solution.
  \label{Fig:Sedov Snapshot}
  }
\end{figure}

In Figure \ref{Fig:Sedov Snapshot} we plot the analytic solution at $t=0.0508$
against the angle average of the numerical solution found by COSMOS using a
uniform $64^3$ grid in the box $[0,1]^3$. At this time the shock front is
located at $r = 0.314$. Since the grid is Cartesian,
in the initial conditions we have attempted to minimize geometrical effects
due to the shape of the grid by using an initial sphere of radius ($\delta r$)
3.5 zones. For each zone containing part of this sphere we weight the initial
pressure according to the fraction of the zone which lies within the sphere
using Monte Carlo averaging. Despite these efforts some small errors of
geometrical origin are still present, particularly in the velocity field,
where some oscillation ($\sim 7\%$) can be seen behind the shock.
The position of the shock itself is accurate to 1--2 zones, which is also the
width of the shock in the numerical solution. However, because the shock
is narrower than one zone in the analytic solution, the density and pressure
in the numerial solution do not reach their maximum analytic values
$\rho_1$ and $p_1$. The density is also slightly underestimated near $r=0.25$,
as is the central pressure.
Given the use of a uniform, relatively coarse Cartesian grid for this
spherically symmetric problem, the position and shape of the shock are
well-determined.

To determine the effect of a nonuniform grid on our Sedov solution,
we solved the same problem on grids with different degrees of nonuniformity.
(The grid in question is static,
not an adaptive grid designed to track the explosion more accurately.)
In these runs the $x$-axis is nonuniform, and the $y$ and $z$ axes are uniform.
Along the $x$-axis the innermost $N_{\rm inner}$ zones are uniform with
spacing $\Delta$. Outside this region the zones increase in width by
$\eta$ per zone toward the outer edges. Each axis uses 64 zones; the
$y$ and $z$ axes each enclose the range $[0,1]$, and for the $x$-axis we fix
$N_{\rm inner}$ at 16 and $\Delta$ at 1/64, allowing $\eta$ to take on
several values
between 0 and 20\%. (Hence the box size in $x$ also varies.) For each run the
explosion is thus allowed to develop up to a certain point within the same
uniformly gridded region; thereafter part of the shock front expands into
a region of nonuniform gridding.

\begin{figure}
\epsscale{1.02}
\plotone{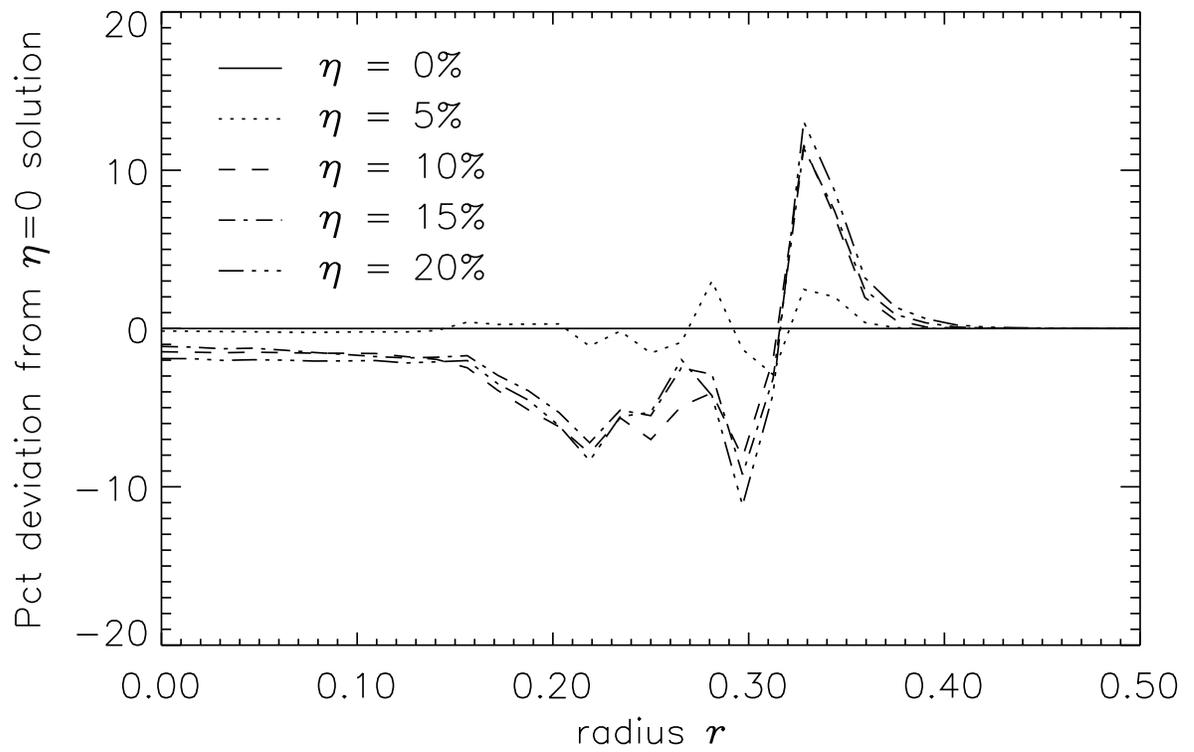}
\caption{
  Percentage deviation from the
  uniform-grid case of the angle-averaged solution to the Sedov
  explosion problem on grids of varying nonuniformity.
  \label{Fig:Sedov Asymmetry Plot}
  }
\end{figure}

Figure \ref{Fig:Sedov Asymmetry Plot} shows, as a function of radius, the
percentage change in the angle-averaged density at $t=0.0508$ for several
values of $\eta$ up to 20\%. Each density profile has been interpolated
to the same uniform grid. Because the shock is only partly resolved
even on the uniform grid at $t=0.0508$, the increased zone width for
large $\eta$ forces the shock to be substantially broadened, leading to
an overestimate of the density ahead of the shock and an underestimate behind
it. The average density is also underestimated by a constant 1--2\% well behind
the shock for $\eta \ge 10\%$ because the broadened shock does not
sufficiently compress the ambient medium. For $\eta=5\%$ the magnitude of
these effects is much smaller.

As the explosion proceeds in the analytic solution, the width of the shock
front increases, and its amplitude decreases. If the zone size increases much
faster than the rate of increase of the width of the shock, the numerical
solution will continue to degrade as the shock propagates outward.
For this problem the value of $\eta$ at which the zone size begins to
increase faster than the shock width appears to lie between 5\% and 10\%.
In general, the amount of nonuniformity to use must be balanced against the
need to resolve fine flow structures near the edge of the grid. If the zone
size is permitted to be too large, the zone-averaged density, pressure, and
velocity may be correct, but the position and size of flow structures will be
poorly determined, and their dynamical effects (such as heating and compression)
on the surrounding gas will also be in error.

\subsubsection{Jeans instability problem}

Using the Jeans instability we verified that the gas dynamical terms
involving the gravitational potential are correctly implemented in COSMOS.
To do this we studied the dispersion relation of stable perturbations to a
uniform, self-gravitating medium. The initial conditions
for this problem, which we solved in two dimensions with periodic boundaries,
are, at time $t=t_i$,
\begin{eqnarray}
\nonumber
\rho(x,y,t_i) & = & \rho_0\left[{1 + \delta\cos(\kappa x)\cos(\kappa y)}\right] \\
\nonumber
p(x,y,t_i)    & = & p_0\left[{1 + \gamma\delta\cos(\kappa x)\cos(\kappa y)}\right] \\
v_x(x,y,t_i)  & = & {\delta|\omega|\over \sqrt{2}\kappa}
		\left[{\sin(\kappa x)\cos(\kappa y)\over 1-\delta\cos(\kappa x)\cos(\kappa y)}\right] \\
\label{Eqn:Jeans Initial Conditions}
\nonumber
v_y(x,y,t_i)  & = & {\delta|\omega|\over \sqrt{2}\kappa}
		\left[{\cos(\kappa x)\sin(\kappa y)\over 1-\delta\cos(\kappa x)\cos(\kappa y)}\right]\ ,
\end{eqnarray}
\noindent where the perturbation amplitude $\delta\ll 1$, and where $\omega$ is
defined as follows. The stability of the perturbation is determined by the
relationship between the wavenumber $k=\sqrt{2}\kappa$ and the Jeans wavenumber $k_J$
(Chandrasekhar 1961), where $k_J$ is given by
\begin{equation}
k_J \equiv {\sqrt{4\pi G\rho_0}\over c_0}\ ,
\end{equation}
\noindent and where $c_0$ is the sound speed:
\begin{equation}
c_0 = \sqrt{\gamma p_0\over\rho_0}\ .
\end{equation}
\noindent If $k > k_J$, the perturbation is stable and oscillates with
frequency
\begin{equation}
\omega = \sqrt{c_0^2k^2 - 4\pi G\rho_0}\ ;
\label{Eqn:Jeans Dispersion Relation}
\end{equation}
\noindent otherwise it grows exponentially, with a characteristic timescale
given by $\tau = (i\omega)^{-1}$.

\begin{figure}
\epsscale{1.02}
\plotone{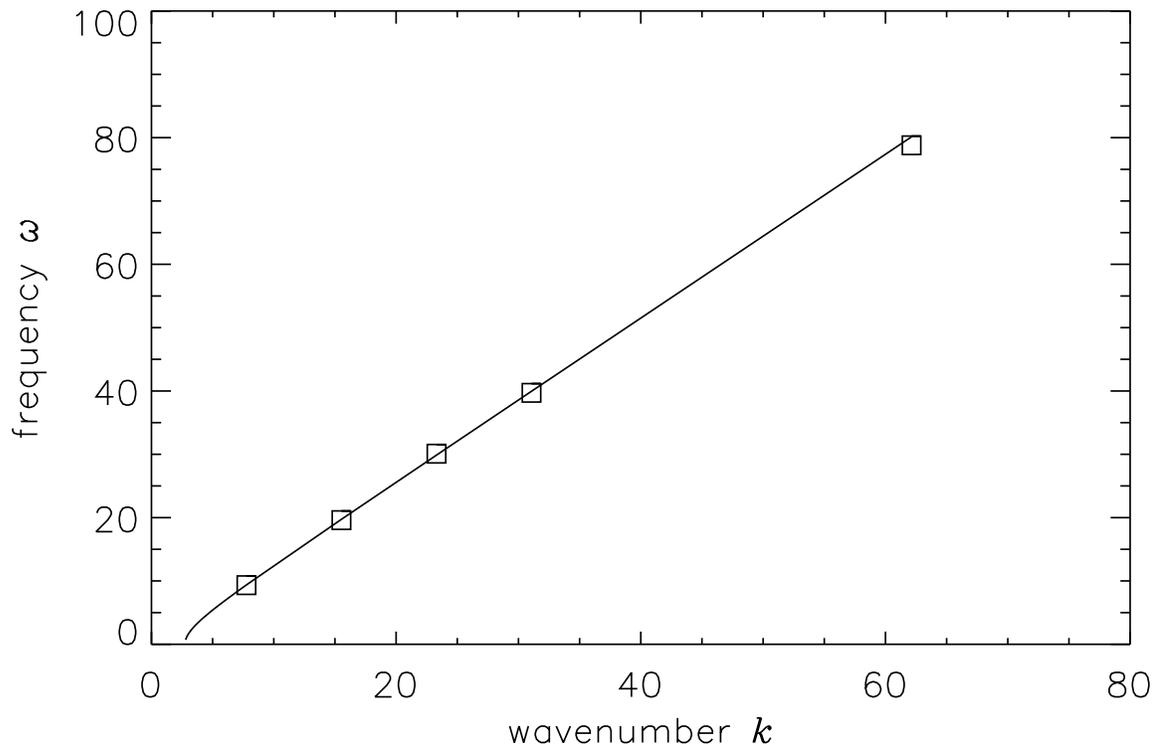}
\caption{
  Test of the Jeans dispersion
  relation for gravitationally stable perturbations. Squares indicate
  oscillation frequencies taken from the test calculations; the solid line
  indicates the expected relation.
  \label{Fig:Jeans Dispersion Plot}
  }
\end{figure}

We checked the dispersion relation (\ref{Eqn:Jeans Dispersion Relation}) for
stable perturbations with $\gamma=5/3$ by fixing $\rho_0$ and $p_0$ and
performing several runs with different $k$. We followed each case for roughly
ten oscillation periods using a uniform $64^2$ grid in the box
$[0,\sqrt{\pi\gamma}/2]^2$ with units in which $\rho_0 = p_0 = G = 1$.
(The box size is chosen so that $k_J$ is smaller than the smallest nonzero
wavenumber which can be resolved on the grid. This prevents numerical errors
at wavenumbers less than $k_J$ from being amplified by the physical Jeans
instability.)
We then computed the oscillation frequency in each case by measuring
the time interval required for the density at the center of the grid to
undergo exactly ten oscillations. The resulting dispersion relation is
compared to equation (\ref{Eqn:Jeans Dispersion Relation}) in Figure
\ref{Fig:Jeans Dispersion Plot}. It can be seen from this plot that our
code correctly reproduces equation (\ref{Eqn:Jeans Dispersion Relation}).
At the highest wavenumber ($k = 62.132$), each oscillation is resolved using
only about 9 cells, and the average timestep (which depends on $c_0$,
$\Delta x$, and $\Delta y$, and has nothing to do with $k$) turns out to be
about one-fifth of an oscillation. Hence the frequency determined from the
numerical solution for this value of $k$ is somewhat more poorly determined
than for the other runs. Overall, however, the frequencies are correct to
about 1\%.

\subsubsection{Zel'dovich pancake}
\label{Sec:Gas pancake}

The cosmological pancake problem (Zel'dovich 1970) provides a good simultaneous
test of the self-gravity and cosmological expansion code. Analytic
solutions well into the nonlinear regime are available for both
$N$-body and hydrodynamical codes
(Anninos \& Norman 1994), permitting an assessment of the code's accuracy.
After caustic formation the problem provides a stringent test of the code's
ability to track thin, poorly resolved features and strong shocks using
most of the basic physics needed for cosmological problems. Also, as pancakes
represent single-mode density perturbations, coding this test problem is
useful as a basis for creating more complex initial conditions.

We set the initial conditions for the pancake problem in the
linear regime using the
analytic solution given by Anninos and Norman (1994). In a universe with
$\Omega_0=1$ at redshift $z$, a perturbation
of wavenumber $k$ which collapses to a caustic at redshift $z_c<z$ has
comoving density and velocity given by
\begin{eqnarray}
\label{Eqn:pancake soln 1}
\rho(x_e;z) & = & \bar{\rho}\left[1 + {1+z_c\over1+z}\cos\left(kx_\ell\right)
	\right]^{-1}\\
\nonumber
v(x_e;z)    & = & -H_0(1+z)^{1/2}(1+z_c){\sin kx_\ell\over kx_\ell}\ ,
\end{eqnarray}
\noindent where $\bar{\rho}$ is the comoving mean density.
Here $x_e$ is the distance of a point from the pancake midplane,
and $x_\ell$ is the corresponding Lagrangian coordinate, found by iteratively
solving
\begin{equation}
x_e = x_\ell - {1+z_c\over 1+z}{\sin kx_\ell\over k}\ .
\end{equation}
\noindent The temperature solution is determined from the density under
the assumption that the gas is adiabatic with ratio of specific heats
$\gamma$:
\begin{equation}
\label{Eqn:pancake soln 2}
T(x_e;z) = (1+z)^2\bar{T}_{\rm fid}\left[\left({1+z_{\rm fid}\over 1+z}
	\right)^3
	{\rho(x_e;z_{\rm fid})\over\rho(x_e;z)}\right]^{\gamma-1}\ .
\end{equation}
\noindent The mean temperature $\bar{T}_{\rm fid}$ is specified at a
redshift $z_{\rm fid}$.

At caustic formation ($z=z_c$), planar shock waves form on either side
of the pancake midplane and begin to propagate outward. A small region
at the midplane is left unshocked. Immediately behind the shocks, the
comoving density and temperature vary approximately as
\begin{eqnarray}
\rho(x_e;z) &\approx&
		\bar{\rho}{18\over\left(kx_\ell\right)^2}
		{\left(1+z_c\right)^3\over\left(1+z\right)^3}\\
\nonumber
T(x_e;z) &\approx&
		{\mu H_0^2\over6k_Bk^2}\left(1+z_c\right)
		\left(1+z\right)^2\left(kx_\ell\right)^2\ .
\end{eqnarray}
\noindent At the midplane, which undergoes adiabatic compression, the
comoving density and temperature are approximately
\begin{eqnarray}
\rho_{\rm center} &\approx&
		\bar{\rho}\left[{1+z_{\rm fid}\over 1+z}\right]^3
		\left[{3H_0^2\mu\over k_B\bar{T}_{\rm fid}k^2}
		{\left(1+z_c\right)^4\left(1+z\right)^3\over
		1+z_{\rm fid}}\right]^{1/\gamma}\\
\nonumber
T_{\rm center} &\approx&
		{3H_0^2\mu\over k_Bk^2} \left(1+z\right)^2
		\left(1+z_c\right)^4{\bar{\rho}\over\rho_{\rm center}}\ .
\end{eqnarray}

To test the convergence of the code, we performed several different 
one-dimensional pancake test runs with fixed perturbation
wavelength and caustic redshift and varying spatial resolution.
Each run used $k=2\pi/(10{\rm\ Mpc})$, $z_c=5$, and an initial redshift
$z_i=50$. In each run we also fixed
$z_{\rm fid}=200$, $\bar{T}_{\rm fid}=550$~K, $\gamma=5/3$,
$\Omega_0=1$, and $H_0=50$~km~s$^{-1}$~Mpc$^{-1}$. We assumed a 75\% H/25\% He
composition for the purpose of computing the pressure using the temperature.
Using the analytic solution (eqs.~[\ref{Eqn:pancake soln 1}] --
[\ref{Eqn:pancake soln 2}]), we computed L1 error norms for density
and temperature in each run at $z\approx 7$, in the mildly nonlinear
phase of the collapse. We define the L1 error norm as in equation
(\ref{Eqn:L1 error}) below.
The results appear in
Figure \ref{Fig:pancake conv}. The density converges slowly, with the
error varying approximately as $\Delta x^{0.6}$. The temperature
converges more rapidly, with the error varying as $\Delta x^{3.5}$ for
$\Delta x>\lambda/128$ and then as $\Delta x^{1.3}$ for smaller $\Delta x$.
The roughly linear asymptotic convergence for the temperature is consistent
with other work, for example that of Bryan et al.\ (1995), but the
density convergence rate is somewhat slower. Nevertheless, the absolute
error of a few percent for $\Delta x\sim\lambda/100$---$\lambda/200$ is
consistent with their results for both variables.

\begin{figure}
\epsscale{1.02}
\plotone{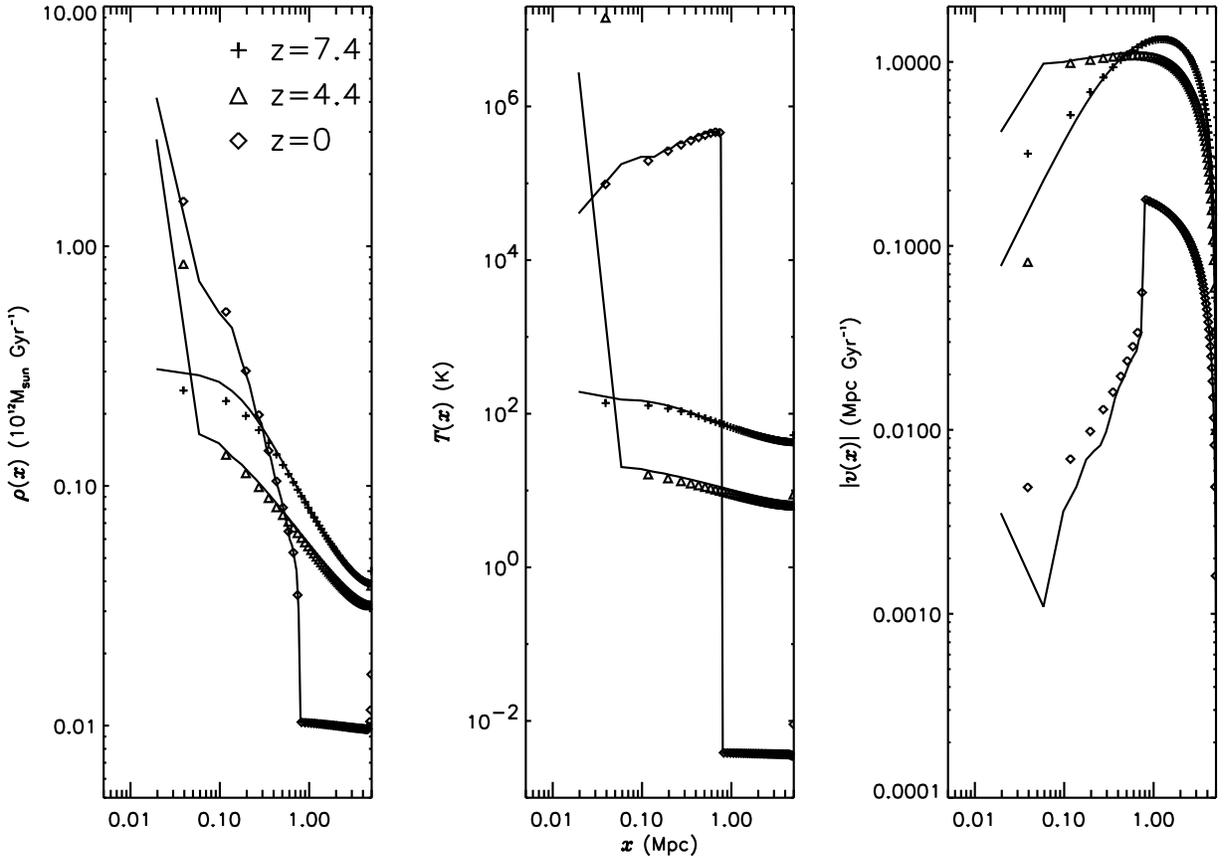}
\caption{
  Comparison of the $256^2$ gas-only pancake test at three different
  redshifts with the 256-zone 1D solution. Density, temperature,
  and velocity amplitude are shown. Symbols represent values interpolated
  from the 2D calculation; lines represent values taken from the 1D
  calculation.
  \label{Fig:pancake comp}
  }
\end{figure}

To examine the performance of COSMOS on this problem in multidimensions,
we performed a run with the same initial conditions using a doubly
periodic $256^2$ grid. The pancake midplane was inclined $45^\circ$ to the
$x$ axis, and the zone spacing
was chosen so that two wavelengths fit into the box diagonal.
In Figure \ref{Fig:pancake comp} the results of this run are compared to
the 256-zone 1D solution at three different redshifts corresponding to
the mildly nonlinear regime ($z=7.4$), the time immediately following caustic
formation ($z=4.4$), and a time well into the nonlinear, post-caustic evolution
($z=0$). The profiles for the $256^2$ run are derived by interpolating
along the grid diagonal. In this run only 128 zones fit within a perturbation
wavelength, so the spatial resolution is one-half that of the 1D 256-zone run.
With the exception of the innermost 1--2 zones, the 2D results are typically
within a few percent of the 1D results. Differences of note include a
20\% underdensity
in the 2D case at $z=7.4$ and a significant overestimate in the central
temperature at $z=4.4$. In the latter case the caustic is unresolved by the
mesh, and only the innermost zone is hot. The velocity in the innermost zone is
different by as much as a factor of two at all three redshifts.


\subsection{Tests of the Poisson solver}

\subsubsection{Miyamoto-Nagai potential}

We tested the potential solver with isolated boundaries using the
Miyamoto-Nagai potential (Miyamoto \& Nagai 1975). This is an
axisymmetric, flattened potential designed to mimic the light
distribution of a disk
galaxy. It is given in cylindrical coordinates by
\begin{equation}
\phi_{\rm MN} (r,z) \equiv - {GM\over\sqrt{r^2 +
					   \left(a+\sqrt{z^2+b^2}\right)^2}}\ ,
\label{Eqn:Miyamoto-Nagai Potential}
\end{equation}
\noindent where $M$ is the total mass, and the ratio of the axis parameters
$b/a$ determines how flattened the potential is. As $b/a\longrightarrow 0$,
$\phi_{\rm MN}$ tends toward the Kuzmin disk potential, and as
$b/a\longrightarrow\infty$, $\phi_{\rm MN}$ tends toward the spherically
symmetric Plummer potential. Thus for different $b/a$ values $\phi_{\rm MN}$
contains different contributions from high-order multipole moments, making
this a good test of our moment-based isolated boundary solver.
The density function corresponding to this potential is straightforward to
derive using the Poisson equation; it is
\begin{equation}
\rho_{\rm MN}(r,z) = \left({b^2M\over 4\pi}\right)
		     {ar^2 + \left(a+3\sqrt{z^2+b^2}\right)
			     \left(a+\sqrt{z^2+b^2}\right)^2 \over
		      \left[r^2+\left(a+\sqrt{z^2+b^2}\right)^2\right]^{5/2}
		      \left(z^2+b^2\right)^{3/2}}\ .
\label{Eqn:Miyamoto-Nagai Density}
\end{equation}

Using this density function we computed Miyamoto-Nagai potentials
with the isolated Poisson solver in COSMOS for $b/a = 0.1$, $1$, and $10$ using 
a $128^2\times 64$ grid in boxes containing 98--99\% of the total mass.
The innermost one-half of the zones in each dimension were uniform, and the
remainder increased in width by 5\% per zone toward the outside edges.
We used units in which $G = M = a = 1$.

\begin{figure}
\plotone{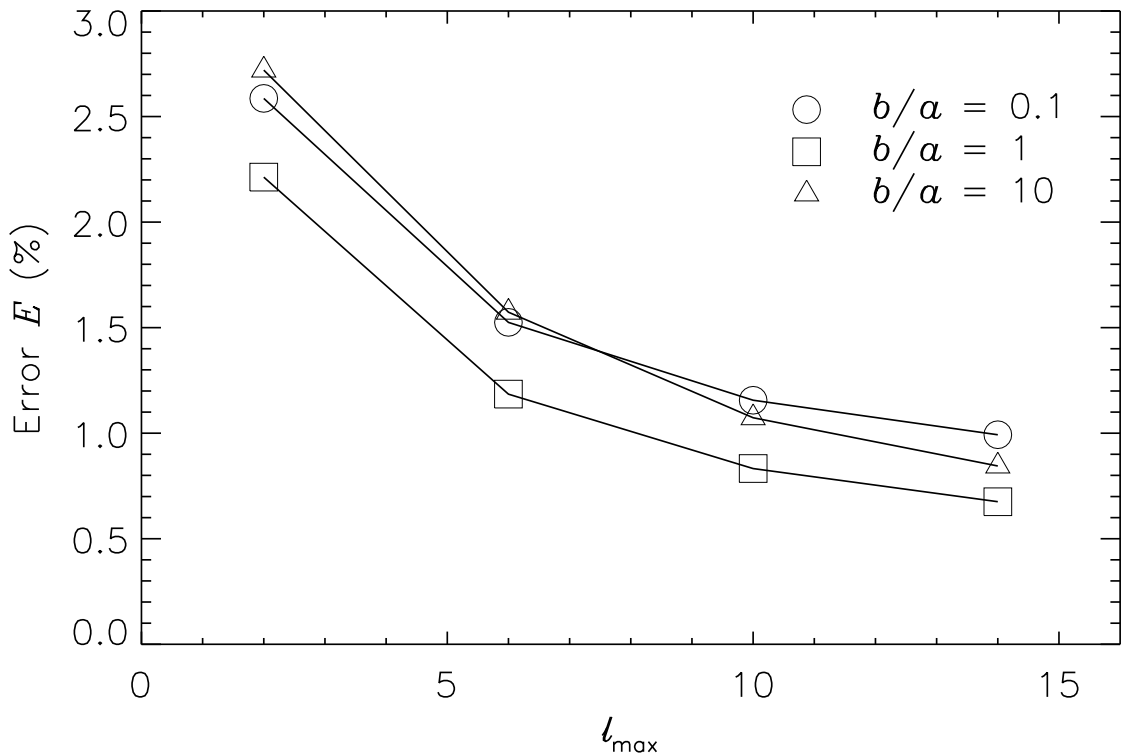}
\caption{
  Average error in Miyamoto-Nagai
  potential calculations for different flatness ratios $b/a$, plotted as a
  function of the largest multipole moment $\ell_{\rm max}$ used in the
  computation of isolated boundaries.
  \label{Fig:Miyamoto-Nagai Error Plot}
  }
\end{figure}

In our treatment of isolated boundaries we compute boundary values for the
image potential using a truncated multipole expansion of the image mass
distribution. To verify that this expansion was implemented correctly, we
performed runs with four different values of $\ell_{\rm max}$, the largest
multipole moment. In Figure \ref{Fig:Miyamoto-Nagai Error Plot} we plot
average errors for our
computed potentials as a function of $\ell_{\rm max}$ for our three values of
$b/a$. We define the error $E(\ell_{\rm max},b/a)$ as
\begin{equation}
\label{Eqn:L1 error}
E(\ell_{\rm max},b/a) \equiv {100\over V}
			     \sum_{ijk} \left |{\phi_{ijk}-\phi_{{\rm MN},ijk}
					\over \phi_{{\rm MN},ijk}}\right|
					\Delta V_{ijk}\ ,
\end{equation}
\noindent where the sum is taken over all of the cells in the grid, and the
total volume $V$ of the grid is the sum of cell volumes $\Delta V_{ijk}$.
Figure \ref{Fig:Miyamoto-Nagai Error Plot} shows that including only the
first three multipole terms already gives a fairly
good estimate of the potential; the maximum percentage error in any cell for
$\ell_{\rm max}=2$ is between 6.5\% and 7.5\%, and the plotted averages range
from 2.2\% to 2.7\%. Increasing the value of
$\ell_{\rm max}$ brings the average and maximum errors down substantially,
showing that the isolated boundary solver is including higher-order moments
properly. The curves begin to level off
at values less than 1\% above $\ell_{\rm max}=10$; this level of error is
consistent with the amount of mass which lay outside the grid and hence
was excluded from each calculation. The maximum error at $\ell_{\rm max}=14$
is about 1.5\% in all three cases.

For other problems the multipole content of the density field may differ, so
these test results do not show that 1\% errors will be achieved 
with $\ell_{\rm max} \approx 10$ for any arbitrarily chosen density field.
However, this will be true even for quite
flattened density fields. Errors caused by underestimating $\ell_{\rm max}$
will be largest near the boundaries of the grid as long as most of the mass
is near the center.

\subsubsection{Particle-mesh force resolution test}

Because of the particle-mesh force-smoothing procedure, dark matter
particles experience an effective potential that deviates from the
Newtonian $1/r$ dependence at small interparticle separations $r$.
This deviation is greatest on length scales less than or comparable to
the zone spacing $\Delta r$. In addition, the introduction of a grid breaks
the rotational symmetry of the
equations of motion, so interparticle forces are not isotropic for
$r\lsim\Delta r$. When performing simulations in which small-scale
structure (relative to the grid in use) appears, it is important to have
an estimate of the magnitude of these errors and the value of
$r/\Delta r$ at which they become important.

To quantify these effects, we performed a test in which we chose 10,000
randomly placed pairs of particles on a $32^3$ grid, computed forces
for each pair using the multigrid solver and particle-mesh code from COSMOS,
and tabulated the resulting forces as functions of the particle positions.
The first particle in each pair was chosen to lie within one zone of the
center of the grid, while the second particle was chosen to lie between
0.1 and 10 zones from the first, with a random position angle relative to
the first.
In this test each particle had unit mass, and the value of $G$ was unity.
For the $i$th particle of the $j$th pair
we computed radial ($f_{ij,r}$) and azimuthal ($f_{ij,\theta}$)
force components relative
to the line connecting the particle to its partner; then for the pair we
computed the average of the magnitudes of the
radial and the azimuthal components for the two particles:
\begin{equation}
\bar{f}_{j,X}  \equiv  {1\over2}\left(\left|f_{1j,X}\right| +
				    \left|f_{2j,X}\right|\right)
\end{equation}
where $X=r,\theta$.
We collected the particle pairs in radial bins of logarithmic width 0.05,
then computed mean and RMS deviations for the binned forces.

\begin{figure}
\epsscale{0.85}
\plotone{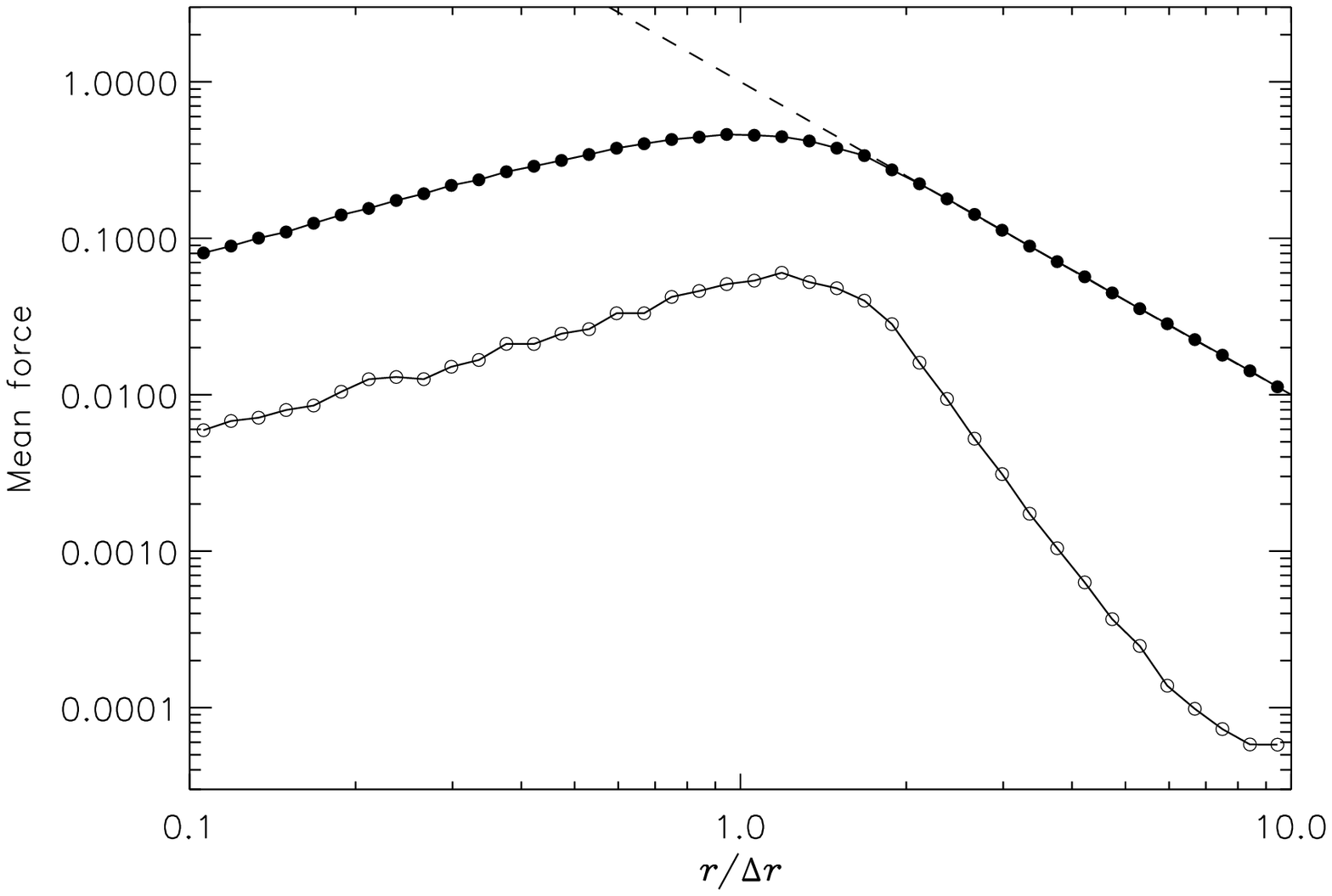}
\plotone{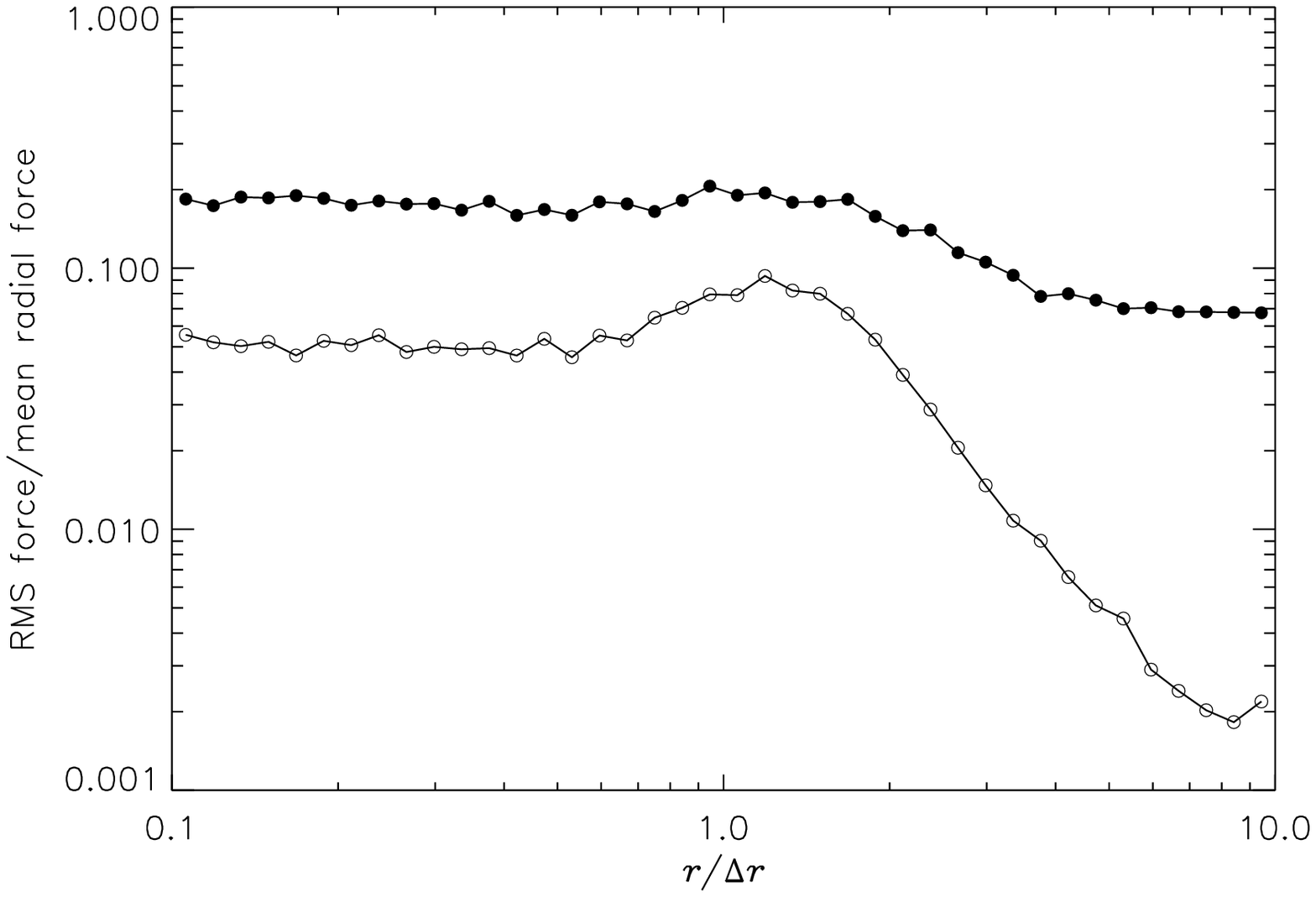}
\caption{
  Results of the PM force resolution test. (a) Mean binned radial (filled
  circles) and azimuthal (open circles) components of the interparticle
  force for randomly positioned particle pairs. The dashed line indicates
  the expected $1/r^2$ force.
  Here $r$ is expressed in units of the zone spacing $\Delta r$.
  (b) RMS deviations for the binned radial and azimuthal components, expressed
  as fractions of the mean binned radial force.
  \label{Fig:force test}
  }
\end{figure}

In Figure \ref{Fig:force test} we show the results as functions of
radius. Overall the results are comparable to those obtained by
Efstathiou et al.\ (1985): our effective particle force resolution is
about two zones. For $r=\Delta r$ the mean radial force is 50\% of the
expected value, and the force anisotropy (ratio of mean azimuthal to mean
radial force) is about 13\%.
For particle separations smaller than this, the mean radial force is
proportional to $r$, and the force anisotropy is roughly constant at around 7\%.
As $r\rightarrow\infty$ the azimuthal force drops as $r^{-3}$, so the
force anisotropy drops approximately as $r^{-1}$. At $r\approx 9$ the
azimuthal force turns upward slightly; this illustrates the effect of placing
the second particle of some pairs within 4--5 zones of the grid boundary.
We discuss this effect in more detail below.

For particle separations less than one zone,
the RMS deviations in the radial and azimuthal force components are each
roughly constant at 20\% and 5\%, respectively, of the mean radial
force. For $r>2\Delta r$
the radial RMS deviation drops to about 7\% of the radial mean, while the
azimuthal RMS deviation decreases almost as fast as the mean azimuthal force,
becoming less than 0.2\% of the radial mean at $r=10\Delta r$.
The azimuthal RMS deviation also shows a slight upturn for particles close
to the boundary.

The deviations from a $1/r^2$ force law at large $r$ result from a
self-force felt only by particles close to the edge of the grid.
As Hockney and Eastwood (1988) show, self-forces arise in PM schemes
when either the density assignment and force interpolation operators
differ or the effective Green's function ${\bf G}_{ijklmn}$ fails
to satisfy the criterion
\begin{equation}
\label{Eqn:Green criterion}
{\bf G}_{ijklmn} = -{\bf G}_{lmnijk}\ .
\end{equation}
The effective Green's function is related to the zone-averaged
gravitational field ${\bf g}_{ijk}$ by
\begin{equation}
{\bf g}_{ijk} = \sum_{lmn}
	{\bf G}_{ijklmn}\rho^{\rm PM}_{lmn}\Delta V_{lmn}\ ,
\end{equation}
where $\rho^{\rm PM}_{lmn}$ is the density assigned to the
grid by the particle-mesh scheme, $\Delta V_{lmn}$ is the volume of a
single grid zone, and the sum is taken over all zones.
Because we use the CIC operator both to assign densities to the grid and to
interpolate forces to particle positions, any self-forces are
due to errors in the potential which cause ${\bf G}$ to violate equation
(\ref{Eqn:Green criterion}). In our case, errors in ${\bf G}$ arise because
we use a finite multipole expansion to obtain boundary conditions for
isolated problems.

\begin{figure}
\epsscale{1.02}
\plotone{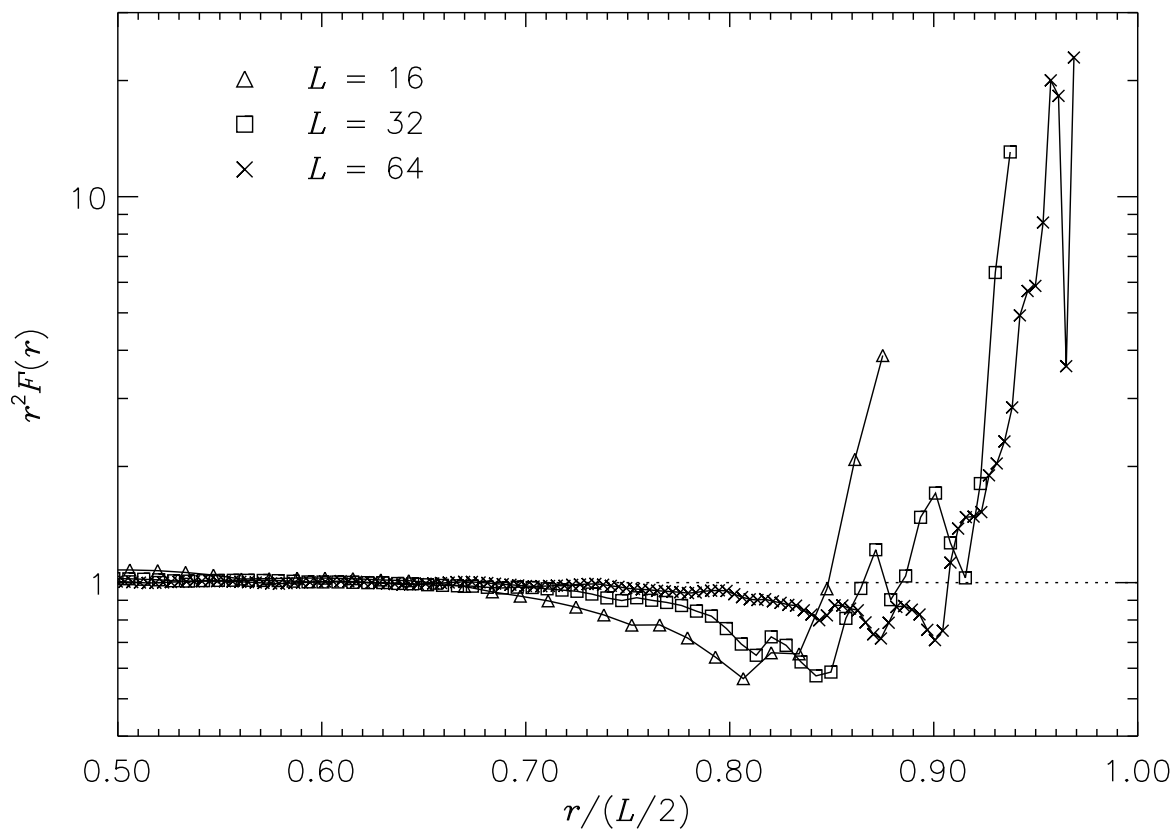}
\caption{
  Results of the PM force resolution test with a single pair of particles,
  where the first member of each pair is placed at the center of the grid
  and the second member is placed at increasing distances from the center
  along the line connecting the center to one of the grid boundaries.
  The plot shows the ratio of the force on the second particle to the
  expected value as a function of the ratio of the particle separation $r$ to
  the box size $L$.
  \label{Fig:force test 2}
  }
\end{figure}

We have examined this effect by performing another test
in which one particle is placed at the center of the grid and the second
particle is placed at increasing distances from the center along one
coordinate axis. By varying the grid size, we can use this test to
determine how close to the edge of the grid a particle can be before it
experiences a significant self-force.
For this test we use $\ell_{\rm max}=10$ in the multigrid solver.
Figure \ref{Fig:force test 2} compares the
results on grids with $16^3$, $32^3$, and $64^3$ zones.
In each case the force on the second particle begins to deviate from the
expected value when the second particle is about five zones from the edge of
the grid.

Note that the two-particle
potential is more susceptible to this effect than the Miyamoto-Nagai test
potential because of the
relative importance in this case of mass close to the edge of the grid.
The effect is important only for small numbers of particles close to the
edge of the grid; it is not important for particles near the boundary
orbiting in the potential of a large group of particles near the center,
and it is not present when periodic boundary conditions are used.



\subsection{$N$-body tests}

\subsubsection{Time integration accuracy}

To account for the varying timestep size, COSMOS implements 
equation (\ref{Eqn:timestep}) to update the dark matter positions
and velocities. To test this algorithm, we set up two particles
in a simple circular orbit and monitored the accuracy of the orbit
as we varied the timestep $\Delta t$ and the zone size $\Delta x$.
In addition to testing
the algorithm, this problem enables us to determine the regions in
$(\Delta x,\Delta t)$ space that are appropriate for
larger, more interesting problems. 

\begin{figure}
\epsscale{0.9}
\plotone{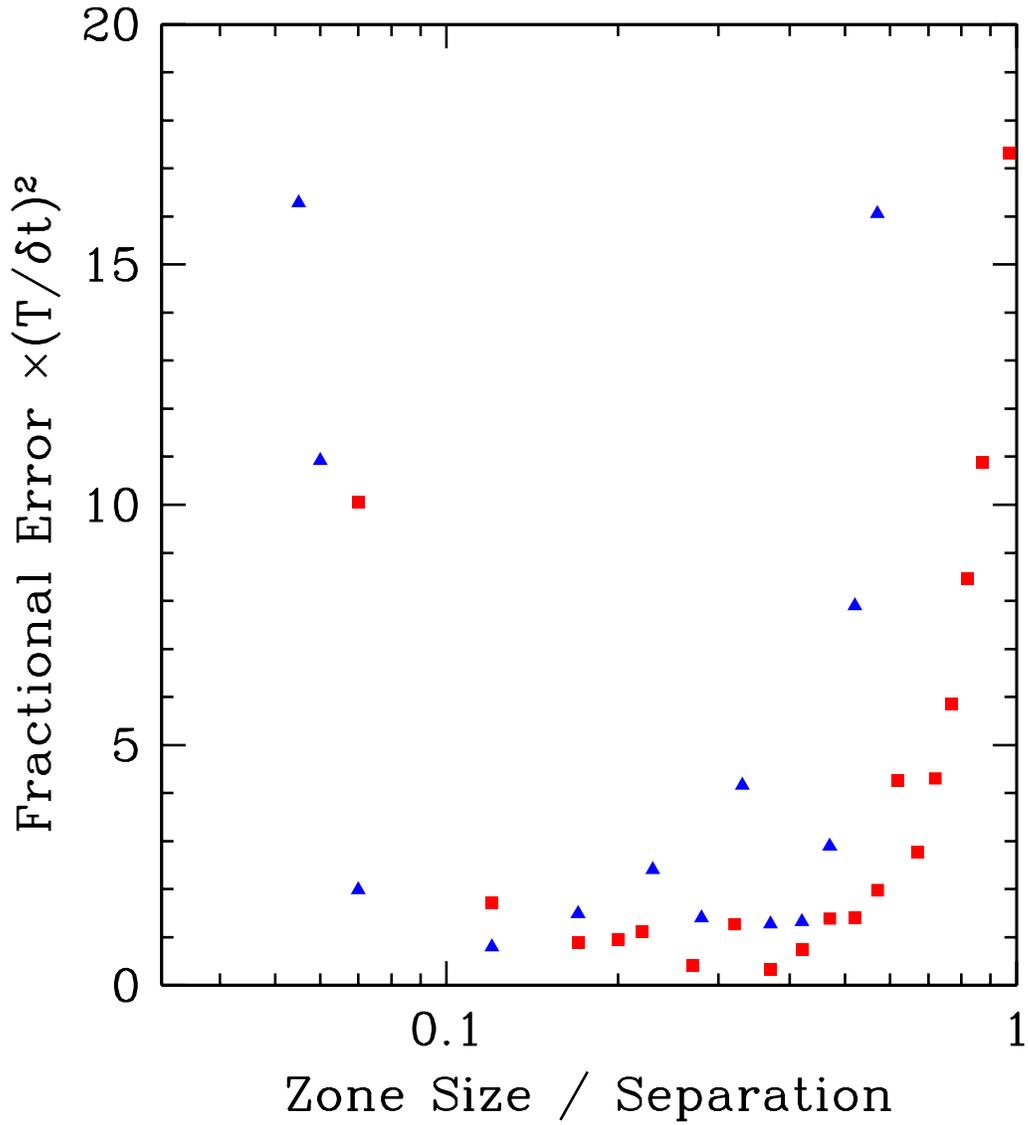}
\caption{
  RMS error in the circular orbit of two particles as a function of the ratio of
  the zone spacing $\Delta x$ to the (fixed) interparticle separation $r$.
  Choices of timestep $\Delta t$ corresponding to 0.015 orbital period
  (triangles) and 0.046 orbital period (squares) are shown.
  \label{Fig:dm_orbit}
  }
\end{figure}

Figure \ref{Fig:dm_orbit} shows the root mean square (RMS) error
in the interparticle separation after six orbital periods as a
function of $\Delta x$.
For $\Delta x$ larger than the particle
separation, the $1/r^2$ force is calculated
incorrectly due to the mesh smoothing.
The problem of resolving forces
on scales smaller than a zone size is a well-known one;
Figure \ref{Fig:dm_orbit}
simply points out that our force resolution is roughly one or two 
zones.
For very small $\Delta x$, on the other hand, each particle
traverses more than one zone in a single timestep, which again leads to
large errors. To alleviate this second problem, in production runs we
choose $\Delta t$ with the requirement that
no dark matter particle travel more than a fraction $\sigma_{\rm dm}$ of a zone
during a single timestep:
\begin{equation}
\Delta t_{\rm dm} = \sigma_{\rm dm}\min_{\rm particles}\left[
	{\Delta x_i\over v_{{\rm dm},x}^p},
	{\Delta y_j\over v_{{\rm dm},y}^p},
	{\Delta z_k\over v_{{\rm dm},z}^p}\right]\ ,
\end{equation}
where $(i,j,k)$ are the indices of the zone containing the $p$th particle.
Figure \ref{Fig:dm_orbit} shows that
setting this fraction equal to one is an acceptable choice.
The two groups of points in this plot correspond to different ratios of
the timestep to the orbital period. If we denote this ratio by $f$, then
we require
\begin{equation}
{\Delta x\over2r} > {\pi f\over\sigma}
\end{equation}
for accuracy. The squares indicate runs with $f=0.046$, and the triangles
indicate runs with $f=0.015$. If we choose $\sigma_{\rm dm}=1$, then
the two groups of runs require $\Delta x/2r > 0.14$ and $0.047$
respectively. For the first group, runs with $\Delta x/2r > 0.1$ are
observed to be
relatively accurate, while for the second group, the error is low for
$\Delta x/2r>0.06$.
In the regime where the zone size is not too big or too small relative to the
timestep, we see that the 
cumulative error does indeed scale as $\Delta t^2$: the algorithm 
of equation (\ref{Eqn:timestep}) is accurate to second order.

\subsubsection{Spherical collapse problem}

A well-known cosmological problem with an analytic solution
is the case of a spherical overdensity (e.~g., Padmanabhan 1993).
The solution to this problem is best described in terms of the radius
$R(t)$ which encloses a mass $M$ at time $t$:
\begin{equation}
R(t;M) = R(t_i;M) \sin^2\left[ {\theta\over 2} \right]
	 { 1 + \delta_i \over \delta_i}\ .
\label{Eqn:Shell}
\end{equation}
Here $t_i$ and $\delta_i$ denote the time and fractional overdensity
at which the simulation
starts. The parameter $\theta$ is a timelike parameter; in a flat,
matter-dominated universe, the
redshift can be expressed in terms of $\theta$ as 
\begin{equation}
1 + z = (1 + z_i) \left[
		1 + {3\over 4} { 1 + \delta_i \over \delta_i^{3/2}} 
		\bigg( \theta - \sin\theta -\theta_i + \sin\theta_i \bigg)
	\right]^{-2/3}.
\end{equation}
Equation (\ref{Eqn:Shell}) holds until the radius reaches a maximum value
at $\theta = \pi$, after which time virialization occurs.

\begin{figure}
\epsscale{0.9}
\plotone{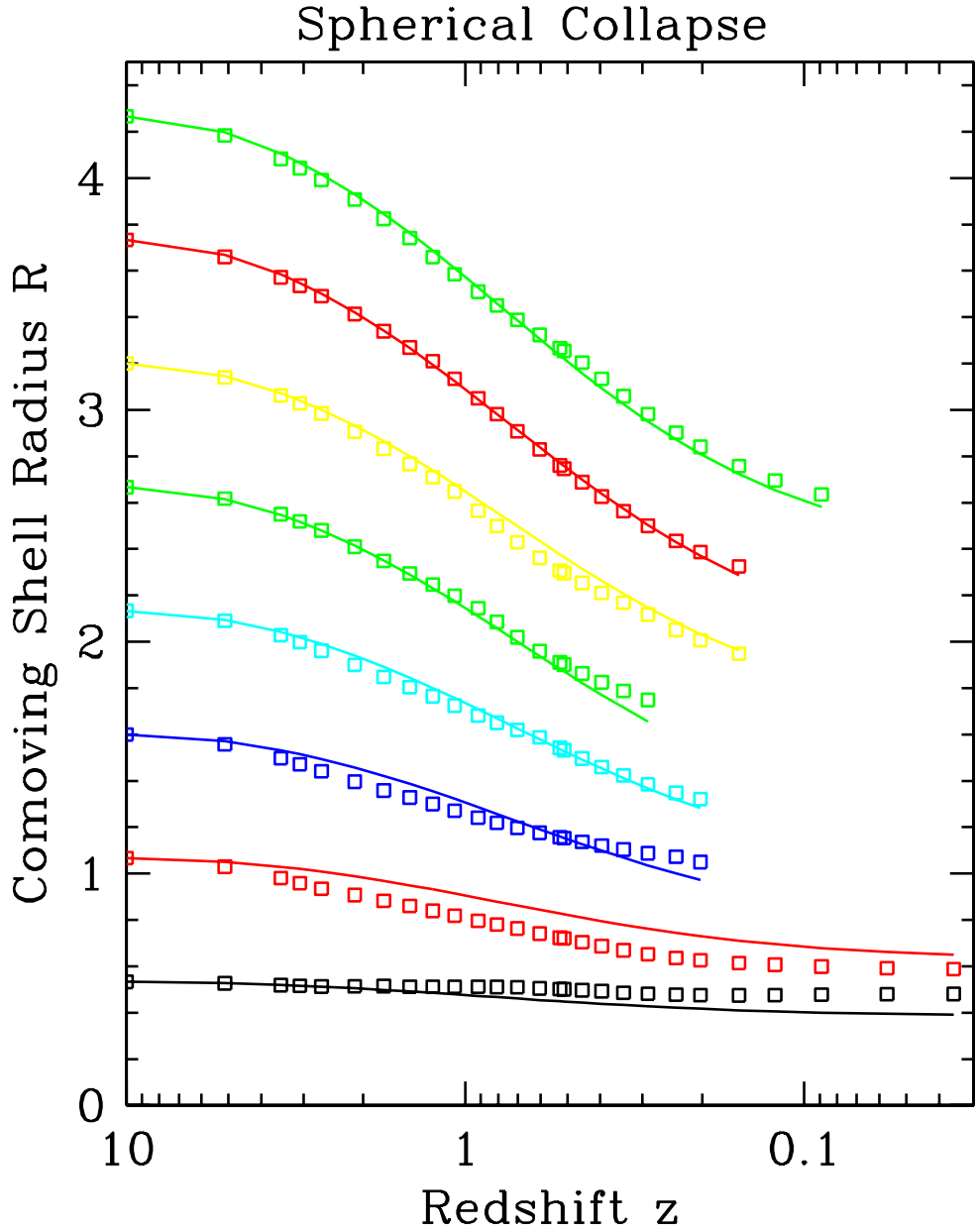}
\caption{
  Time evolution of the radius of different mass shells in the dark-matter-only
  spherical collapse test. Units are such that the zone spacing is unity.
  Points are results from a $32^3$ simulation; solid lines are the analytic
  solution, plotted until the time of maximum physical radius.
  \label{Fig:dm_spherical}
  }
\end{figure}

On a $32^3$ grid with $32^3$ particles we set up spherically symmetric
initial conditions.
In units in which the cell size is unity, within a radius of five units from the
center of the grid we constructed a uniform overdensity with $\delta_i = 0.3$.
In an annulus ranging
from $r = 5$ to $r=8$ we placed an underdensity so that the 
average density within
$r=8$ was equal to the background density. Figure \ref{Fig:dm_spherical}
shows the time evolution of several (comoving) radii, compared with the
analytic result of equation (\ref{Eqn:Shell}).
The larger shells track the analytic
solution well, verifying that all factors of $a(t)$ are correctly implemented
in the dark matter code.
As expected, the small shells do worse: the force on these shells comes 
predominantly from particles concentrated at a distance of one to two zones.
The deviations from equation (\ref{Eqn:Shell}) for these shells are
consistent with the underestimate of the gravitational force introduced by
the particle-mesh smoothing at this distance.

\subsubsection{One-dimensional Plane Wave}

The Zel'dovich pancake (Section \ref{Sec:Gas pancake}) also serves as a
test of the $N$-body code.
To initialize this problem, we place particles at the zone centers
of a $16^3$-zone grid, then perturb their positions slightly
in the $x$-direction by amounts
\begin{equation}
\delta x_i \propto \cos(2\pi k x_i /L)
\end{equation} 
with $x_i$ the unperturbed positions and $k$ an integer.
Small-scale perturbations correspond to large $k$.
We consider only perturbations which vary
on scales larger than a zone width, ie.\ those with $ k << N_x$. If the
particles' peculiar velocities $\delta v_{x,i}$ are also proportional to
$\cos(2\pi k x_i /L)$, then in the linear regime (assuming a flat universe)
both $\delta x_i$ and $\delta v_{x,i}$ should grow with time as the scale factor
$a(t)$.

\begin{figure}
\epsscale{0.9}
\plotone{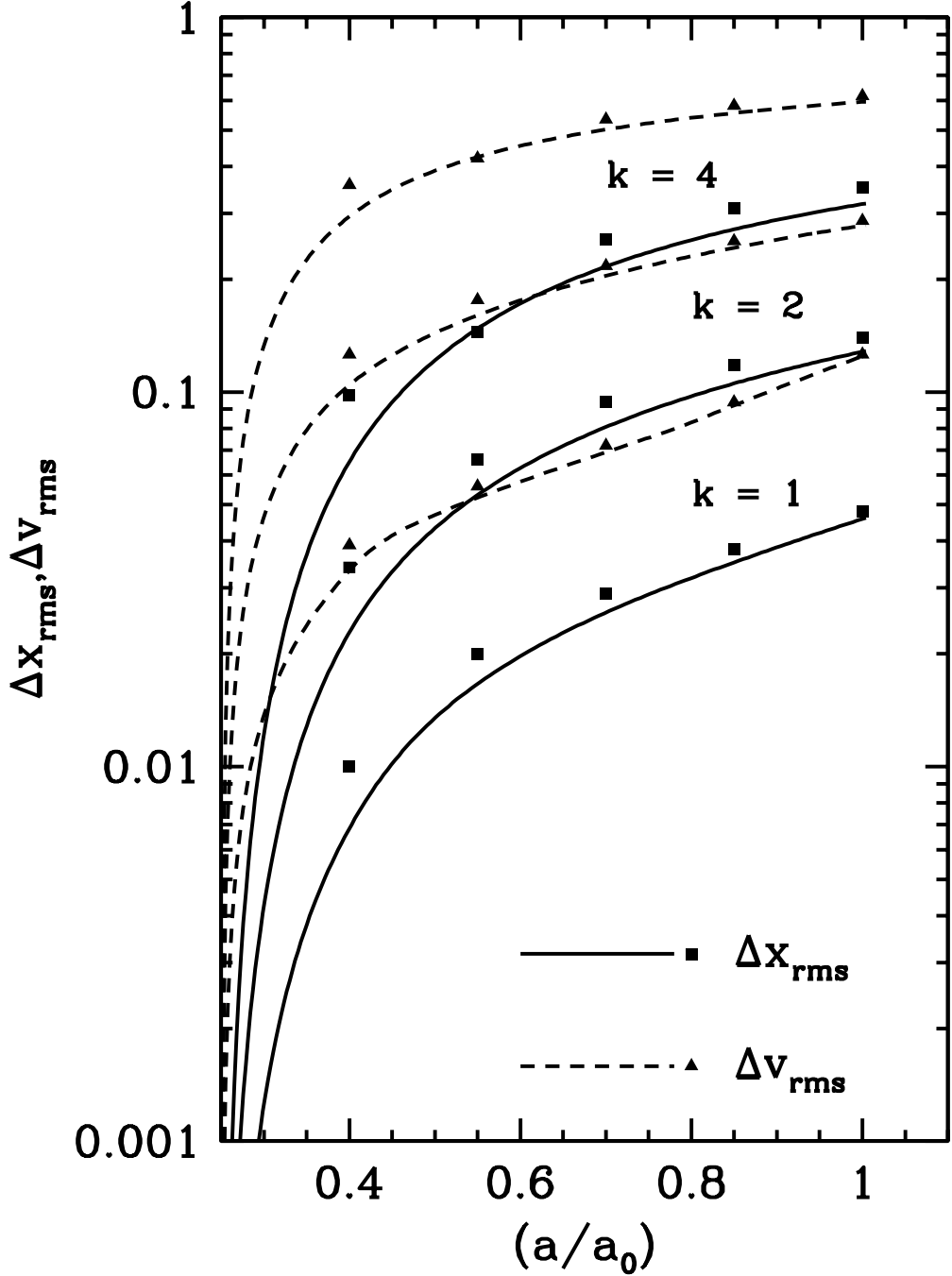}
\caption{
  RMS errors in position ($\Delta x_{\rm RMS}$) and velocity ($\Delta v_{\rm
  RMS}$) for the dark-matter-only plane wave problem in the linear regime.
  Lines represent results from simulations using a $16^3$ grid and $16^3$
  particles. Results for three different values of the wavenumber $k$ are shown.
  Points represent values from equivalent runs by
  Efstathiou \etal\ (1985), who used an FFT Poisson solver instead of the
  multigrid method used here.
  \label{Fig:cdeltarms}
  }
\end{figure}

To determine how accurately their code
could follow such perturbations, Efstathiou et~al.\ (1985) tabulated
the RMS deviation of position and velocity from the analytic solution
for particles on a $16^3$ mesh. Since we solve Poisson's
equation differently than the standard PM code, it is useful to compare our
results with theirs. Figure \ref{Fig:cdeltarms} shows the RMS errors in position
and velocity for linear perturbations with several different values of the
wavenumber $k$. The multigrid solver and variable-timestep integrator
used in COSMOS produce results quite similar to (but slightly better than) the
FFT/fixed-timestep technique used by Efstathiou et~al.\ (1985).



\section{Conclusion}
\label{Sec:Conclusion}

In this paper we have described COSMOS, a new hybrid code for solving
cosmological
problems involving gas and collisionless matter, including self-gravity,
cosmological expansion, and radiative cooling. COSMOS solves the inviscid
fluid equations using the piecewise-parabolic method on a static, nonuniform
structured grid. The code treats collisionless matter using the cloud-in-cell
variant of the particle-mesh method, and it computes the gravitational
potential using a linear full multigrid solver. COSMOS supports both periodic
boundaries, suitable for large-scale structure problems, and isolated
boundaries, suitable for simulations of isolated systems. This code has
already been used to study mergers between clusters of galaxies by
Ricker (1998) and Ricker \& Sarazin (2000).

Several new features of COSMOS distinguish it from other existing PM-PPM codes.
In particular, the multigrid Poisson solver can determine the gravitational
potential of isolated matter distributions and properly
takes into account the finite-volume
discretization required by PPM. All components of the code are constructed
to work with a nonuniform mesh, preserving second-order spatial differences.
The PPM code uses vacuum boundary conditions
for isolated problems, preventing inflows when appropriate.
The PM code uses a second-order variable-timestep time integration scheme.
In this paper we have discussed the equations solved by COSMOS and described
the algorithms used, with emphasis on these features.
We have also reported on results from a
suite of standard test problems, demonstrating that COSMOS works as
expected.

In closing we note some implementation details.
COSMOS is designed to run on parallel computers using the
Parallel Virtual Machine (PVM) message-passing library.
A version using the Message-Passing Interface (MPI) is currently
under development. Also, COSMOS features a modular design
which simplifies the addition of new physics and the configuration of the
code for different types of problems.
The modular design of COSMOS has served as the basis for FLASH, an
adaptive mesh-refinement PPM code currently under development at the
University of Chicago ASCI Flash Center. Details of the FLASH code design will
appear in a forthcoming paper (Fryxell et al.\ 1999).



\acknowledgments

PMR would like to thank Craig Sarazin for
advice and support during the completion of this work
and Bruce Fryxell, Kevin Olson, and Peter MacNeice for
useful and stimulating discussions during its initial stages.
He also acknowledges support from NASA through
a GSRP Fellowship (NGT-51322) and NAG 5-3057, as well as support
from the DOE ASCI Flash Center under contract B341495.
SD is supported by NASA under NAG 5-7092 and the DOE.
DQL acknowledges support from NASA under NAG 5-2868.
The calculations reported here were performed at the Pittsburgh
Supercomputing Center, the San Diego Supercomputer Center, and
the Texas Advanced Computing Center.



\newpage

\newcommand{\and}{\ \&}
\newcommand{\CPC}{Comp.\ Phys.\ Comm.}
\newcommand{\JCP}{J.\ Comp.\ Phys.}
\newcommand{\MNRAS}{MNRAS}
\newcommand{\ApJ}{ApJ}
\newcommand{\ApJS}{ApJS}
\newcommand{\AJ}{AJ}
\renewcommand{\AA}{A\&A}
\newcommand{\PASJ}{PASJ}



\clearpage

\clearpage

\clearpage

\clearpage

\clearpage

\clearpage

\clearpage

\begin{figure}
\epsscale{0.85}
\plotone{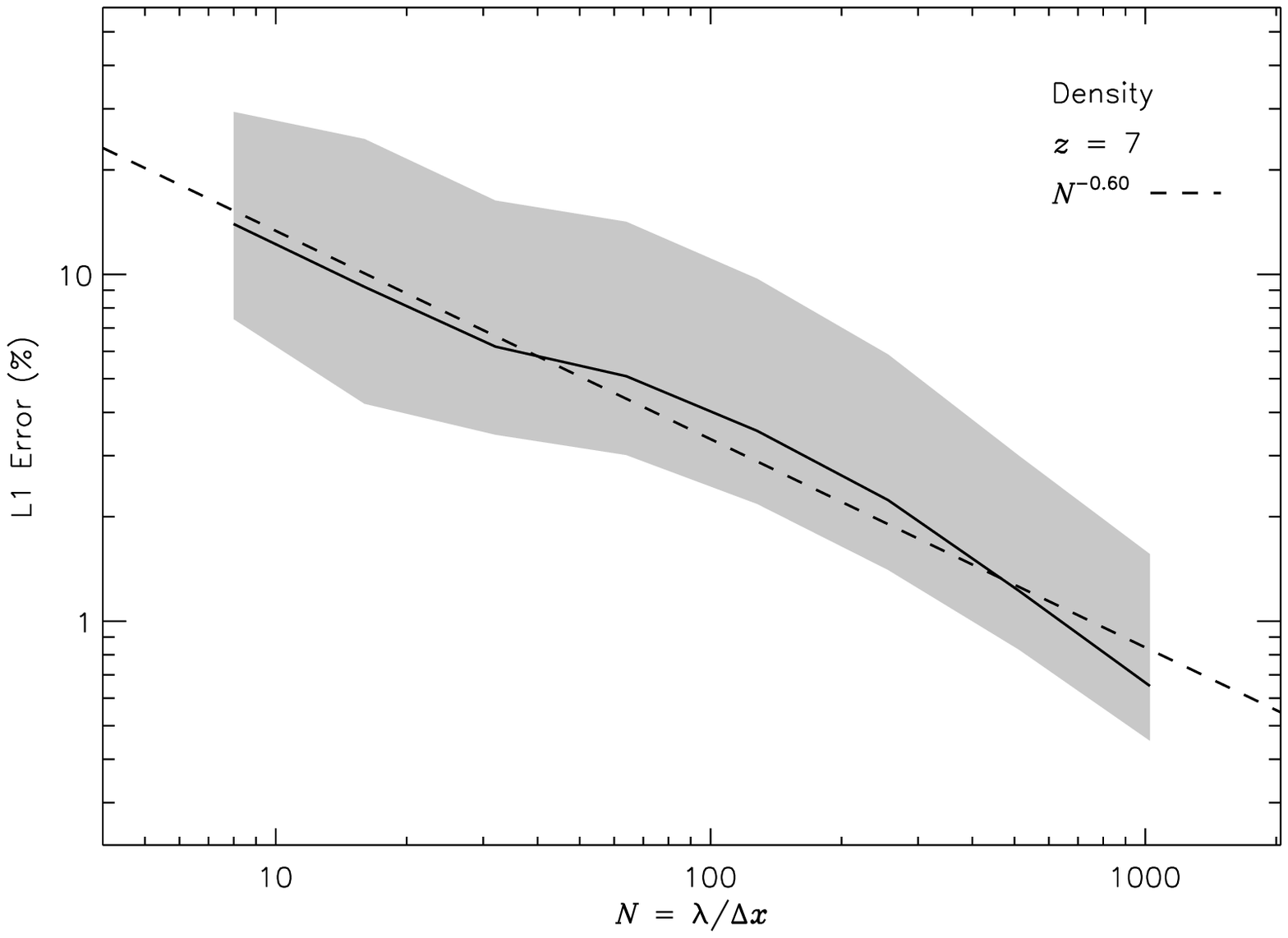}
\plotone{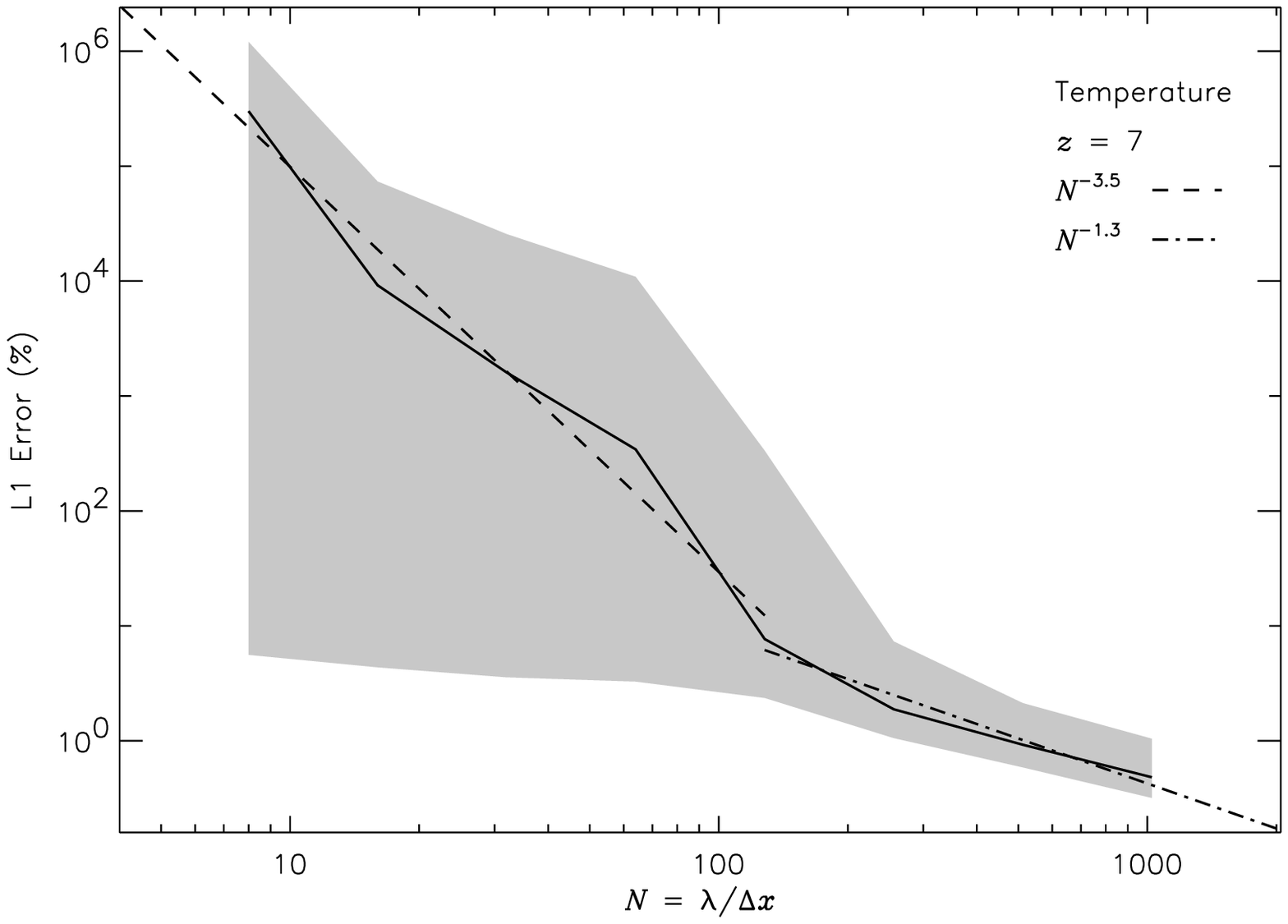}
\caption{
  One-dimensional convergence results for the gas-only Zel'dovich pancake
  problem, as functions of the number of zones $N$, for fixed
  perturbation wavelength $\lambda$.
  Solid lines indicate the L1 error norm (eq.~[\ref{Eqn:L1 error}]),
  while shaded areas indicate RMS deviations of the error from the L1 norm.
  Dashed lines indicate power-law relations for comparison.
  \label{Fig:pancake conv}
  }
\end{figure}

\clearpage

\clearpage

\clearpage

\clearpage

\clearpage

\clearpage

\clearpage

\end{document}